\newif\iftechreport
\techreporttrue
\RequirePackage{etoolbox}
\ifdef{\iftechreport}{}{%
  \newif\iftechreport
  \techreportfalse
}

\documentclass[conference]{IEEEtran}

\newif\ifanonymous
\anonymousfalse

\usepackage{cite}
\usepackage{textcomp}
\usepackage{xcolor}
\usepackage{xspace}
\usepackage{comment}
\usepackage{float}
\def\BibTeX{{\rm B\kern-.05em{\sc i\kern-.025em b}\kern-.08em
    T\kern-.1667em\lower.7ex\hbox{E}\kern-.125emX}}

\usepackage{amsmath}
\ifcsname Bbbk\endcsname\else
    \usepackage{amssymb}
\fi
\usepackage{amsfonts}
\usepackage{mathrsfs}
\usepackage{latexsym}
\usepackage{bm} 

\usepackage[normalem]{ulem}
\usepackage{graphicx}
\usepackage[caption=false,font=footnotesize]{subfig}
\makeatletter
\renewcommand{\@IEEEfigurecaptionsepspace}{\vskip2pt\relax\nointerlineskip}
\renewcommand{\@IEEEtablecaptionsepspace}{\vskip2pt\relax\nointerlineskip}
\makeatother

\usepackage{amsthm}
\usepackage{thmtools,thm-restate}
\usepackage[ruled,vlined,linesnumbered]{algorithm2e}
\SetKwProg{Fn}{Function}{}{}
\SetVlineSkip{0pt}
\SetAlCapNameFnt{\small}
\SetAlCapFnt{\small}
\SetAlFnt{\small}
\makeatletter
\patchcmd\algocf@Vline{\vrule}{\vrule \kern-0.4pt}{}{}
\patchcmd\algocf@Vsline{\vrule}{\vrule \kern-0.4pt}{}{}
\patchcmd{\@algocf@start}{-1.5em}{0pt}{}{}
\makeatother

\usepackage{multicol}
\usepackage{multirow}
\usepackage{booktabs}

\usepackage[hidelinks]{hyperref}
\usepackage[capitalise,noabbrev]{cleveref}
\crefname{appendix}{appendix}{appendices}
\Crefname{appendix}{Appendix}{Appendices}

\makeatletter
\patchcmd{\subsubsection}
  {\@startsection{subsubsection}{3}{\parindent}}
  {\@startsection{subsubsection}{3}{\z@}}
  {}{\PackageError{paper}{Cannot set subsubsection headings flush left}{Check the IEEEtran subsubsection definition.}}
\makeatother

\renewcommand{\paragraph}[1]{\par\noindent\hspace*{\parindent}\textbf{#1}\enspace\ignorespaces}

\apptocmd\normalsize{%
   \setlength{\abovedisplayskip}{2pt}
   \setlength{\belowdisplayskip}{2pt}
   \setlength{\abovedisplayshortskip}{2pt}
   \setlength{\belowdisplayshortskip}{2pt}
}{}{}

\theoremstyle{plain}
\newtheorem{theorem}{Theorem}[section]

\newtheorem{lemma}[theorem]{Lemma}

\theoremstyle{definition}

\newtheorem{definition}[theorem]{Definition}
\theoremstyle{plain}
\pretocmd{\appendices}{%
  \crefalias{section}{appendix}%
}{}{\PackageError{paper}{Cannot configure appendix references}{Check the IEEEtran appendices definition.}}

\usepackage{paralist}
\usepackage[inline]{enumitem}
\setlist{noitemsep,partopsep=0pt,topsep=0pt}
\setlist[itemize,1]{leftmargin=2em}

\newcommand{\blue}[1]{#1}

\newcommand{\ie}{\textit{i.e.}\xspace}
\newcommand{\eg}{\textit{e.g.}\xspace}

\newcommand{\indname}{MeST\xspace}
\newcommand{\schemename}{VIDA\xspace}

\newcommand{\concat}{\mathbin{|}}
\newcommand{\D}{\mathcal{D}}

\newcommand{\funcname}[1]{\textnormal{\texttt{#1}}}

\begin{document}
\title{Authenticated Approximate Nearest Neighbor Search in Outsourced Vector Databases}

\ifanonymous
\author{\IEEEauthorblockN{Anonymous Author(s)}}
\else
\author{%
\IEEEauthorblockN{\fontsize{11}{13}\selectfont Yuanhang Zhou, Xinyi Zhang, Cheng Xu, Jianliang Xu}
\IEEEauthorblockA{\textit{Department of Computer Science, Hong Kong Baptist University, Hong Kong SAR}\\
{\small\ttfamily\{csyhzhou, csxyzhang, chengxu, xujl\}@comp.hkbu.edu.hk}}}
\fi

\maketitle

\begin{abstract}

Approximate Nearest Neighbor (ANN) search has become a fundamental query primitive in modern vector databases for managing large-scale embedding data. To support growing vector collections, organizations increasingly outsource database management and query processing to third-party clouds. %
However, untrusted service providers may return suboptimal or fabricated results, posing a serious threat to ANN search integrity. Existing authenticated ANN approaches face substantial scalability challenges. They typically require transmitting high-dimensional vectors for distance verification and rely on topology-agnostic authentication structures, resulting in excessive communication and verification overhead. In this paper, we propose VIDA, a novel and efficient scheme for authenticated graph-based ANN search. VIDA replaces raw vector transmission with compact vector commitments and employs an aggregated Inner Product Argument to efficiently verify multiple distance computations. Furthermore, we introduce MeST, a topology-aware authenticated data structure that aligns authentication paths with graph traversal, significantly reducing structural proof overhead. Additionally, we design a quantization-aware authentication strategy that exploits distance bounds for pruning, reducing the number of exact distances requiring cryptographic authentication. Extensive experiments demonstrate that the proposed schemes substantially improve the efficiency of authenticated ANN search, achieving up to a 6.6$\times$ speedup in end-to-end latency and reducing data transfer time by up to 99.2\% compared with state-of-the-art methods.

\end{abstract}

\section{Introduction}
Driven by the rapid advancement of deep learning, massive volumes of unstructured data, such as text, images, and videos, are routinely transformed into high-dimensional vector embeddings to capture their underlying semantic relationships. To efficiently manage and query these immense datasets, specialized vector databases (\eg, Milvus~\cite{Milvus}, Qdrant~\cite{qdrant}, and Weaviate~\cite{weaviate}) have emerged and are now widely deployed. Serving as the core retrieval engine for these systems, ANN search has established itself as a fundamental primitive in the era of artificial intelligence. Leveraging this capability, vector databases empower a wide range of data-intensive applications, including retrieval-augmented generation~\cite{rag}, recommendation systems~\cite{recommender}, and semantic search~\cite{Dense_Passage_Retrieval}. To support ANN search on a massive scale, modern vector databases rely on advanced indexing structures, such as hashing~\cite{lsh}, inverted files~\cite{pq}, and proximity graphs~\cite{HNSW, nsg}. In this paper, we focus on proximity-graph-based (or simply graph-based) indexing methods, as they achieve the best trade-off between search efficiency and recall~\cite{vector_database_survey}. 

With the exponential growth in dataset sizes and the continuous increase in vector dimensionality, maintaining massive vector databases locally has become prohibitively expensive for many data owners. Consequently, there is a growing trend to outsource vector data management to third-party cloud platforms, leveraging their scalable storage and computational capabilities to support ANN queries~\cite{song2026vector2,BlendHouse,upreti2025cost}. Despite the benefits of outsourcing, this paradigm introduces critical trust concerns regarding the execution integrity of the search service. For instance, a rational but dishonest platform might terminate the query process early and return suboptimal results to reduce computational overhead, thereby compromising the accuracy of the results. While authenticated query processing over outsourced data has been extensively investigated for traditional relational databases \cite{integridb,vsql,zksql}, it remains largely under-explored for high-dimensional vector databases.

To ensure the integrity of ANN search in outsourced vector databases, a common approach is to employ traditional authenticated data structures (ADS), such as the Merkle Hash Tree (MHT), to build an authenticated graph index~\cite{annproof_baseline1,graph_baseline2}. Under this approach, upon executing an ANN query, the cloud service provider constructs and returns a verification object (VO) that contains the graph traversal path and all the raw high-dimensional vectors evaluated during the search. The client can then verify the integrity of the query results by locally replaying the search process and recomputing the distances. 
However, existing methods~\cite{annproof_baseline1,graph_baseline2} necessitate the transmission of large volumes of raw high-dimensional vectors in the VO for local distance recomputation. Furthermore, their MHTs are constructed without considering the underlying topology of the graph index, resulting in inefficient authentication paths. Consequently, these limitations lead to an excessively large VO size, incurring high bandwidth costs and resulting in prohibitive communication overhead.

To overcome this challenge, we propose VIDA, a novel and efficient scheme for authenticated graph-based ANN search that effectively addresses the critical bandwidth bottleneck. 
Unlike traditional baselines that require the transmission of raw high-dimensional vectors, VIDA integrates established vector commitments and Inner Product Argument (IPA) into a novel authentication scheme tailored to graph-based ANN search, enabling compact distance verification without transmitting raw vectors.
Furthermore, to address the challenge of verifying a massive number of distance computations during graph traversal, we propose a proof aggregation mechanism that compresses the verification tasks for multiple visited nodes into a single IPA proof, significantly reducing client-side computational overhead. %

\blue{Beyond optimizing data-level transmission, we further propose MeST, a topology-aware ADS that minimizes the structural proof size. MeST is built using a novel distance-aware minimum spanning tree to align authentication paths with graph traversal, thereby reducing the size of Merkle proofs required for verification. Even after these optimizations, the residual verification cost still scales with the number of distance-evaluated candidates. We therefore extend VIDA to support lower-bound-based quantization-aware graph search, in which pruned candidates are justified by compact authenticated quantization payloads, while only the surviving candidates require exact-distance authentication via the aggregated IPA proof. Extensive experiments demonstrate that our methods outperform baseline methods, achieving a substantial speedup in end-to-end latency and reducing data transfer by orders of magnitude in high-recall scenarios.}

Our main contributions are summarized as follows:
\begin{itemize}
    \item %
    We propose VIDA, a novel verifiable graph-based ANN search scheme that integrates vector commitments and IPA to authenticate distances without transmitting raw high-dimensional vectors.

    \item %
    We design a proof aggregation mechanism that compresses multiple inner-product verifications into a single compact proof, greatly reducing client-side computational overhead.
    
    \item %
    We introduce MeST, a topology-aware ADS built via a distance-aware minimum spanning tree, which aligns verification paths with graph traversal to minimize Merkle proof size.

    \item  \blue{We adapt VIDA to quantization-aware graph search, reducing the number of exact distances that require cryptographic authentication.}

    \item We provide a formal security analysis and conduct extensive experiments on both real-world and synthetic datasets. The results show that our methods achieve up to a \blue{6.6$\times$} speedup in end-to-end latency and a \blue{99.2\%} reduction in data transfer time compared with state-of-the-art methods.

\end{itemize}

The rest of the paper is organized as follows. \Cref{sec:prob_def} introduces our research problem, including the system model and security definitions. \Cref{sec:preliminaries} reviews the preliminaries and presents a strawman solution. We detail the design of \schemename in \Cref{sec:VIDA} and the \indname index structure in \Cref{sec:MeST}. \blue{\Cref{sec:qua} further extends \schemename to support quantization-aware authentication.} \Cref{sec:analysis} summarizes the security guarantees, followed by the experimental evaluations in \Cref{sec:evaluation}. Finally, we discuss related work in \Cref{sec:related_work} and conclude the paper in \Cref{sec:conclusion}.

\section{Problem Definition}\label{sec:prob_def}

\textbf{Approximate Nearest Neighbor Search.}
Consider a database $\mathcal{D}$ consisting of a set of data objects, each represented as $o_i = (ID_i, \mathbf{x}_i)$, where $ID_i$ is a unique identifier and $\mathbf{x}_i$ is the embedding vector of the original data payload (\eg, string, graph, or time series). The distance between any two objects is evaluated by a metric function $\delta(\cdot, \cdot)$.
Formally, a $k$-Nearest Neighbor ($k$NN) query $Q = (k, \mathbf{x}_q)$ requests the $k$ nearest neighbors of the query vector $\mathbf{x}_q$. The exact query result set, denoted as $R_{exact} \subseteq \D$, satisfies $|R_{exact}| = k$ and $\delta (\mathbf{x}_q, \mathbf{x}_i) \leq \delta (\mathbf{x}_q, \mathbf{x}_{i'})$ for any $o_i \in R_{exact}$  and any $o_{i'} \in \D \setminus R_{exact}$. 
In practice, exact $k$NN search is computationally prohibitive in high dimensions due to the curse of dimensionality~\cite{HNSW}. This paper therefore focuses on Approximate Nearest Neighbor (ANN) search. By relaxing exactness, ANN algorithms achieve substantially higher query efficiency. Accuracy is measured by the recall rate: $\textsf{Recall} = \frac{|R \cap R_{exact}|}{k}$, where $R$ is the approximate result set returned by the ANN search. The goal of ANN search is to maximize recall while minimizing query latency.

In this paper, we study authenticated ANN search over high-dimensional embedding vectors. We primarily focus on deterministic graph-based ANN algorithms under Euclidean distance, cosine similarity, and inner product, which together cover the majority of real-world ANN queries.

\begin{figure}[t]
    \centering
    \includegraphics[width=0.96\linewidth]{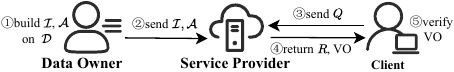}
    \caption{System Overview}
    \label{fig-workflow}
\end{figure}

\noindent
\textbf{System Model.}
Our system, as illustrated in \cref{fig-workflow}, comprises three entities: the Data Owner (DO), the Service Provider (SP), and the Query Client.

\begin{itemize}
\item \textbf{DO:} The DO holds the original database $\mathcal{D}$ and builds an ANN search index $\mathcal{I}$ over $\mathcal{D}$ using a designated deterministic ANN algorithm. To support verifiable query execution, the DO constructs an Authenticated Data Structure (ADS) $\mathcal{A}$ on top of $\mathcal{I}$. The DO then outsources both $\mathcal{I}$ and $\mathcal{A}$ to the SP. To establish trust and ensure the integrity of the ADS, the DO signs the digest of $\mathcal{A}$ and publishes the signature to the client. 
\item \textbf{SP:} Equipped with abundant computational resources, the SP hosts $\mathcal{I}$ and $\mathcal{A}$ and provides query services on behalf of the DO. Upon receiving a query $Q$ from the client, the SP executes the designated ANN algorithm over $\mathcal{I}$ to compute the result $R$ and constructs a verification object (VO) based on $\mathcal{A}$ to attest to its integrity.
\item \textbf{Query Client:} The client is an end-user who issues search queries to the SP. Upon receiving the result $R$ and its corresponding VO, the client verifies the integrity and correctness of $R$ by validating the VO.
\end{itemize}

\noindent
\textbf{Threat Model.}
The main security threat arises because the SP is an untrusted third party that may deviate arbitrarily from the protocol. For example, it may return incorrect results to save computation by skipping neighbors or tampering with the index or vector values. In contrast, we assume that both the DO and the query client are honest and follow the protocol.
To counter this threat, the SP must return a VO along with the query result $R$ to prove execution integrity. At the same time, because exact $k$NN search is prohibitively expensive in high dimensions, we do not require the SP to return the exact result. Therefore, %
we formally define the security of authenticated ANN query processing via security games.

\begin{definition}[Security]
\label{def:security}
We say an authenticated ANN search protocol is \emph{secure} if, for any probabilistic polynomial-time (PPT) adversary \textsf{Adv}, the adversary's success probability in the following experiment is negligible:

\begin{itemize}
    \item \textbf{Setup:} The adversary \textsf{Adv} chooses a database $\mathcal{D}$ and a query $Q$. A trusted party runs the index and ADS generation protocol on~$\mathcal{D}$.
    \item \textbf{Ideal Game ($\textbf{Game}_{\textsf{Ideal}}$):} A trusted party faithfully executes the deterministic ANN algorithm to obtain the ideal result $R_\textsf{Ideal}$. The game outputs $R_\textsf{Ideal}$.
    \item \textbf{Real Game ($\textbf{Game}_{\textsf{Real}}$):} The adversary \textsf{Adv} generates a result $R_\textsf{Real}$ together with a VO. The trusted party runs the verification algorithm $\funcname{Verify}(Q, R_{\textsf{Real}}, \text{VO})$. If it outputs \texttt{reject}, the game outputs $\bot$; otherwise, the game outputs $R_\textsf{Real}$.
\end{itemize}

We say \textsf{Adv} succeeds if it can produce a query result that passes verification but differs from the ideal result $R_\textsf{Ideal}$. Formally:
\begin{equation*}
    \Pr[(\textbf{Game}_{\textsf{Real}} \neq \bot) \land (\textbf{Game}_{\textsf{Real}} \neq \textbf{Game}_{\textsf{Ideal}})] \le \textsf{negl}(\lambda)
\end{equation*}
where $\lambda$ is the security parameter.
\end{definition}

\section{Preliminaries}\label{sec:preliminaries}
In this section, we introduce background on graph-based ANN algorithms and a strawman approach for authenticated ANN query processing. The frequently used notations throughout this paper are summarized in \Cref{table-symbol}.

\begin{table}[t]
	\centering
	\caption{Frequently Used Notations}\label{table-symbol}
    \footnotesize
    \setlength{\tabcolsep}{5pt}
    \renewcommand{\arraystretch}{0.90}
		\begin{tabular}{cc}
			\toprule
			\textbf{Symbol} & \textbf{Description}\\
            \midrule
            
            $\mathcal{D},\mathcal{I},\mathcal{A}$ & vector database, graph-based ANN index, and ADS\\
            $v_i, Adj_i$ & node and adjacency list\\
            $o_i, ID_i, \mathbf{x}_i$ & data object, identifier, and embedding vector\\
            $\mathbf{x}_q, R$ & query vector and result set\\
            $v_{\mathrm{ep}}$, $\delta(\cdot, \cdot)$ & entry point of $\mathcal{I}$, and distance metric function\\
            $H_i$, $H_\mathrm{root}, \sigma$ & node hash, root hash of $\mathcal{A}$, and DO's signature\\
            $C_i, \|\mathbf{x}_i\|$ & vector commitment and norm of $\mathbf{x}_i$ \\
            $IP_i, \rho$ & inner product of $\mathbf{x}_i$ and $\mathbf{x}_q$, and IPA proof\\
            $g_i$ & graph ADS\\
            $V_\mathrm{graph}, V_\mathrm{merkle}$ & set of visited nodes and set of proved nodes in \indname\\
            $V_\mathrm{exa},V_\mathrm{qua}$ & distance-evaluated and quantization-pruned node sets \\
            \bottomrule
		\end{tabular}

\end{table}

\subsection{Graph-based ANN Algorithms}
Graph-based algorithms have been recognized as the most effective category for high-dimensional ANN search. Representative index structures, such as HNSW~\cite{HNSW} and NSG~\cite{nsg}, have demonstrated superior efficiency and accuracy. Consequently, they have been widely adopted in large-scale vector database systems~\cite{Milvus,qdrant,weaviate}. In general, these algorithms operate in two primary phases: offline index construction and online query processing.

During the index construction phase, given a database $\mathcal{D}$, the construction algorithm builds a proximity graph $\mathcal{I}$. Each node $v_i$ in the graph corresponds to a data object $o_i \in \mathcal{D}$, while the edges capture the neighborhood proximity between data points. The index is usually built incrementally by inserting points sequentially, and each new point is connected to a set of selected neighbors subject to an out-degree limit.
During the query processing phase, given an ANN query $Q=(k, \mathbf{x}_q)$, the search algorithm initiates a greedy traversal over the graph $\mathcal{I}$, starting from a designated entry node $v_{\mathrm{ep}}$. Throughout the traversal, a priority queue of candidate nodes is maintained, typically ordered by their distance to the query vector $\mathbf{x}_q$. The algorithm iteratively expands the search frontier by extracting the most promising candidates and evaluating their unvisited neighbors. %
After the traversal terminates according to the specific stopping condition, the top-$k$ closest nodes encountered during the search are returned as the ANN query result $R$.

\subsection{Strawman Solution}\label{sec:strawman}

A strawman solution to enable authenticated graph-based ANN search is to build an MHT~\cite{merkletree} as the underlying ADS over all graph nodes in $\mathcal{I}$. During the query phase, in addition to executing the prescribed ANN search algorithm, the SP is required to generate the corresponding Merkle proofs for all visited graph nodes. These visited nodes, along with their cryptographic proofs, constitute the verification object (VO). Upon receiving the VO, the client first verifies the data integrity of the visited nodes using their Merkle proofs. Subsequently, a subgraph is constructed from these verified nodes, allowing the client to locally replay the ANN search on this subgraph to verify the returned results.

\subsubsection{ADS Construction}

The ADS of the strawman solution is a binary MHT built on top of the graph nodes.
Specifically, each graph node $v_i$ is represented as a leaf node. The leaf node hash is computed as $H_i = \mathbb{H}(ID_i \concat  Adj_i \concat \mathbf{x}_i)$, where $ID_i$ and $\mathbf{x}_i$ are the identifier and raw vector of the corresponding data object, $Adj_i$ is the adjacency list consisting of the identifiers of its neighbors on the index graph, $\mathbb{H}(\cdot)$ denotes a cryptographic hash function, and $\concat$ represents the concatenation operation. By contrast, the non-leaf node hash is computed as $\mathbb{H}(H_l \concat H_r)$, where $H_l$ and $H_r$ are the hashes of the left and right child nodes.
Finally, the DO signs the root hash of the MHT and the identifier of the entry point (\ie, $(H_\mathrm{root}, ID_{\text{ep}} )$) using its private key to generate a digital signature $\sigma$ as the trusted root.

\textit{\textbf{Example.}
Consider a proximity graph as shown in \Cref{fig-graph}. The strawman approach constructs a corresponding MHT, illustrated in \Cref{fig-baseline_index}; \eg, the hash of leaf node $v_1$ is computed as $H_1 = \mathbb{H}(ID_1 \concat  Adj_1 \concat \mathbf{x}_1)$, whereas the hash of the non-leaf node $v_{13}$ is computed as  $H_{13} = \mathbb{H}(H_1 \concat H_2)$. The DO signs $(H_\mathrm{root}, ID_\mathrm{ep} = 1 )$ to generate $\sigma$.
} 

\subsubsection{Verifiable Search}\label{sec:straw_search}
Given a query $Q$, the SP executes the designated graph-based ANN search and records the set of visited candidate nodes $V_\mathrm{graph}$ (\ie, all nodes whose distances to the query vector are evaluated during the traversal, including the entry point ${v_{\mathrm{ep}}}$). It then constructs a VO consisting of: (1) the identifier $ID_i$, raw vector $\mathbf{x}_i$, and graph adjacency list $Adj_i$ for all $v_i \in V_\mathrm{graph}$; (2) the corresponding Merkle proof for these nodes; and (3) the DO's signature $\sigma$. The SP returns the result set $R$ along with the VO. 

To verify the query result, the client first uses the Merkle proof in the VO to recompute the root hash from the received MHT nodes. It then authenticates the reconstructed hash against the one signed by the DO. This validates the integrity of all returned $v_i \in V_\mathrm{graph}$. Next, the client assembles these authenticated nodes to build a local subgraph $\mathcal{I}'$. Finally, the client replays the search over $\mathcal{I}'$ to obtain a local result $R'$. The verification succeeds \textit{iff} $R' = R$ and the local traversal does not try to %
access any unverified node outside~$V_\mathrm{graph}$.

\begin{figure}[t]
    \centering
    \subfloat[A Proximity Graph\label{fig-graph}]{%
        \includegraphics[width=0.47\linewidth]{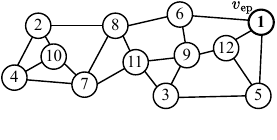}%
    }%
    \hfill
    \subfloat[An Example of Search\label{fig-graph_with_search}]{%
        \includegraphics[width=0.48\linewidth]{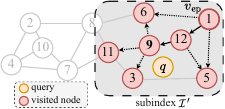}%
    }%
    \caption{An Example of Graph-based ANN search}
    \label{fig-combined}
\end{figure}

\begin{figure}[t]
    \centering
    \includegraphics[width=0.9\linewidth]{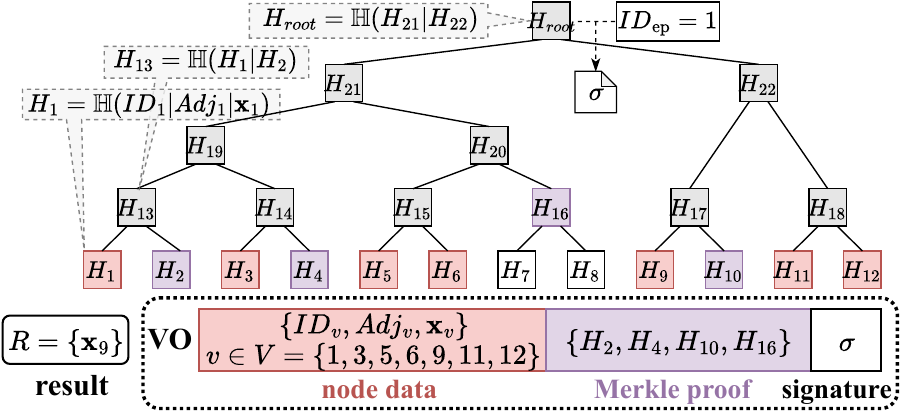}
    \caption{Strawman Index}
    \label{fig-baseline_index}
\end{figure}

\textit{\textbf{Example.}
In \Cref{fig-graph_with_search}, for a 1-NN query $Q$, the traversal starts from entry point $v_1$ and visits the node set $V_\mathrm{graph}=\{v_1, v_3, v_5, v_6, v_9, v_{11}, v_{12}\}$ to find the nearest neighbor $v_9$. Using the MHT in \Cref{fig-baseline_index}, the SP generates a Merkle proof for all nodes in $V_\mathrm{graph}$. To attest to the integrity of a single node (\eg, $v_1$), the standard Merkle proof requires its raw vector along with the sibling hashes $H_2, H_{14}, H_{20}, H_{22}$. When proving multiple nodes simultaneously, redundant hashes can be pruned by aggregating shared paths. For instance, nodes $v_1$ and $v_3$ share the sibling hashes $H_{20}, H_{22}$ when recomputing $H_{19}, H_{21}, H_\mathrm{root}$. The resulting minimal proof for $V_\mathrm{graph}$ is $\{H_2,H_4,H_{10},H_{16}\}$, which must be sent to the client \textbf{alongside the raw high-dimensional vectors of all nodes in $V_\mathrm{graph}$.}}

\textit{For result verification, the client first %
recomputes the leaf hashes $H_1$, $H_3$, $H_5$, $H_6$, $H_9$, $H_{11}$, $H_{12}$ from the received node information. It then iteratively reconstructs the node hashes (\eg, $H_{13}$) by combining the recomputed leaf hashes with the provided sibling hashes until the Merkle root $H_{\mathrm{root}}$ is obtained. The client then checks $\sigma$ against this reconstructed root, thereby verifying the correctness and integrity of the node data.
Subsequently, the client assembles the authenticated nodes into a local subgraph (right part of \Cref{fig-graph_with_search}). It then replays the search for $Q$ over this subgraph from entry point $v_1$ and visits $v_5,v_6,v_{12},v_{9},v_{3},v_{11}$ successively. The verification succeeds if the local result is $R'=\{\mathbf{x}_9\}$ and the traversal only accesses nodes within this subgraph.}

\subsubsection{Limitations} \label{sec:limitation}
The strawman solution suffers from two main limitations that render it impractical for large-scale deployments.

\begin{figure}[t]
    \centering
    \captionsetup[subfloat]{farskip=2pt}
    \subfloat[Breakdown of VO Size]{%
        \includegraphics[width=0.472\linewidth]{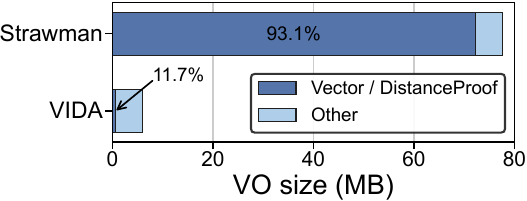}%
        \label{fig:vo_size}%
    }%
    \hfill
    \subfloat[Breakdown of Query Latency]{%
        \includegraphics[width=0.472\linewidth]{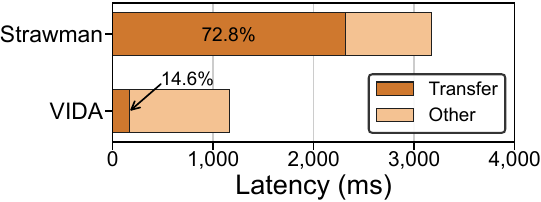}%
        \label{fig:latency}%
    }%
    \caption{Preliminary Evaluation on OpenAI3072}
    \label{fig:motivation}
\end{figure}

\textbf{Prohibitive Communication Overhead.} 
The strawman approach generates an unacceptably large VO, imposing a severe burden on network bandwidth. To enable local search replay, the protocol must transmit the complete raw high-dimensional vector $\mathbf{x}_i$ for every visited node $v_i \in V_\mathrm{graph}$. This bottleneck is especially severe in modern AI applications that use high-dimensional embeddings.
To quantify the overhead, we conducted a preliminary evaluation on the OpenAI3072 dataset~\cite{openai3072} ($d=$ 3072). As illustrated in \Cref{fig:motivation}, transmitting raw vectors constitutes the primary bottleneck of the strawman solution. Specifically, the raw vectors account for 93.1\% of the 77.61~MB VO size. Consequently, data transfer alone consumes 72.8\% of the end-to-end query latency.

\textbf{Topology-Agnostic ADS.} 
The MHT is constructed based on arbitrarily assigned node IDs and remains oblivious to the underlying proximity graph topology. Consequently, semantically close nodes that are frequently co-visited during search (\eg, $v_1,v_9$, and $v_{12}$ in \Cref{fig-graph_with_search}) might be scattered across distant leaves of the MHT and share few common ancestors. This topological misalignment severely limits proof compression, leading to larger proof sizes and higher verification overhead. For instance, sequential nodes $\{v_1,v_2,v_3\}$ share deep common ancestors $H_{19}, H_{21}, H_\mathrm{root}$, enabling compact aggregation, whereas the actual search trajectory $\{v_1,v_9,v_{12}\}$ shares only the root $H_\mathrm{root}$. This structural deficiency %
results in substantially larger %
proofs and inflated verification costs.

\section{\schemename Scheme}\label{sec:VIDA}
In this section, we address the first challenge, namely the prohibitive communication overhead. %
To solve this issue, we propose \schemename, a \underline{V}erifiable \underline{I}nner-product-based \underline{D}istance \underline{A}uthentication scheme that allows the client to authenticate distance computations without receiving the raw vectors.

\subsection{Inner-product-based Distance Authentication}
Preliminary evaluations in \cref{fig:motivation} indicate that the primary bottleneck of the strawman solution is the requirement for the client to receive the complete raw vector $\mathbf{x}_i$ for every visited node $v_i \in V_\mathrm{graph}$ to replay the search process locally.
This overhead is particularly expensive for modern machine learning workloads, where the dimensionality of embeddings often reaches thousands and continues to grow.
Ideally, it would be much preferable if the client could attest to the distance between $\mathbf{x}_q$ and $\mathbf{x}_i$ without receiving $\mathbf{x}_i$ itself. 
To achieve this, we analyze the algebraic structure of common distance metrics used in vector databases:
\begin{itemize}
    \item \emph{Euclidean distance:} $\left(\|\mathbf{x}_i\|^2 + \|\mathbf{x}_q\|^2 - 2\langle \mathbf{x}_i, \mathbf{x}_q \rangle\right)^{\frac{1}{2}}$.
    \item \emph{Cosine similarity:} $\frac{\langle \mathbf{x}_i, \mathbf{x}_q \rangle}{\|\mathbf{x}_i\| \cdot \|\mathbf{x}_q\|}$.
    \item \emph{Inner product:} $\langle \mathbf{x}_i, \mathbf{x}_q \rangle$.
\end{itemize}

A crucial observation is that these metrics depend on three components: the norm of the query vector $\|\mathbf{x}_q\|$, the norm of the data vector $\|\mathbf{x}_i\|$, and the inner product 
$
IP_i = \langle \mathbf{x}_i, \mathbf{x}_q \rangle.
$ 
Since the client provides the query vector $\mathbf{x}_q$, its norm $\|\mathbf{x}_q\|$ can be computed locally. Meanwhile, the DO can precompute the norm of each data vector $\|\mathbf{x}_i\|$ for every object $o_i \in \mathcal{D}$ as part of the ADS. Consequently, the remaining problem reduces to proving the correctness of the inner product $IP_i$. To this end, we propose to leverage the \emph{Inner Product Argument (IPA)}~\cite{linearproof,springproofs}, a cryptographic protocol that allows a prover to convince a verifier of the inner product of two vectors represented by constant-size commitments. The IPA protocol consists of the following PPT algorithms:
\begin{itemize}
\item \funcname{KeyGen}$(1^\lambda) \to pp$: Given a security parameter $1^\lambda$, it generates a public key $pp$.
\item \funcname{Com}$(\mathbf{x}_i, pp) \to C_i$: Given a vector $\mathbf{x}_i$ and the public key $pp$, it generates a constant-size, computationally binding Pedersen vector commitment~\cite{pedersen}.
\item \funcname{ProveIPA}$(\mathbf{x}_1, \mathbf{x}_2, pp) \to (IP, \rho)$: Given vectors $\mathbf{x}_1$, $\mathbf{x}_2$ and the public key $pp$, it computes the inner product $IP = \langle \mathbf{x}_1, \mathbf{x}_2 \rangle$ and the corresponding IPA proof $\rho$.
\item \funcname{VerifyIPA}$(C_1, C_2, IP, \rho, pp) \to \{0, 1\}$: Given the commitments $C_1$, $C_2$, the claimed inner product $IP$, a proof $\rho$, and the public key $pp$, it outputs 1 if and only if $\langle \mathbf{x}_1, \mathbf{x}_2 \rangle = IP \land \funcname{Com}(\mathbf{x}_1) = C_1 \land \funcname{Com}(\mathbf{x}_2) = C_2$.
\end{itemize}
Because the size of a Pedersen commitment is independent of the vector dimensionality $d$, replacing raw vectors with commitments immediately removes the dominant communication overhead associated with transmitting $\mathbf{x}_i$. Furthermore, as we will see later, the IPA enables efficient verification of the inner product computations required for ANN queries and can be extended to support the aggregation of multiple proofs.

\subsection{\schemename Design}

We now describe how \schemename leverages inner-product-based distance authentication to support authenticated ANN search.

\subsubsection{ADS Construction} 
The ADS in \schemename is structured similarly to the strawman solution, employing a binary MHT where each leaf node corresponds to a graph node $v_i$.
However, the leaf node hash is computed as $H_i = \mathbb{H}(ID_i \concat Adj_i \concat C_i \concat \|\mathbf{x}_i\|)$, where $C_i$ and $\|\mathbf{x}_i\|$ are the Pedersen commitment and the norm of $\mathbf{x}_i$, respectively. Note that including the raw vector $\mathbf{x}_i$ in the digest is unnecessary, as the commitment $C_i$ itself serves as a cryptographic binding to the data object. The remainder of the ADS structure and its construction algorithm are identical to those described in \cref{sec:strawman}.

\subsubsection{Verifiable Search}\label{sec:VIDA-search}
Given a query $Q$, the SP executes the designated graph-based ANN search over the index and records the visited node set $V_\mathrm{graph}$. 
Unlike the strawman solution, for each visited node $v_i \in V_\mathrm{graph}$, the VO includes the commitment $C_i$ and norm $\|\mathbf{x}_i\|$ instead of the raw vector $\mathbf{x}_i$, while retaining the identifier $ID_i$ and adjacency list $Adj_i$.
Meanwhile, the SP generates an IPA proof $\rho_i$ using $\funcname{ProveIPA}(\mathbf{x}_i, \mathbf{x}_q, pp)$ for each visited node $v_i$ to attest to the inner product $IP_i = \langle \mathbf{x}_i, \mathbf{x}_q \rangle$. 
These authenticated inner products, combined with the commitments and norms, replace raw-vector transmission and enable the client to recompute distances without accessing $\mathbf{x}_i$, reducing communication overhead. Finally, the SP generates the Merkle proof for $V_\mathrm{graph}$ and returns the result set $R$ along with the VO.
In summary, the VO in VIDA comprises:
\begin{itemize}
    \item the identifier $ID_i$, the adjacency list $Adj_i$, the commitment $C_i$, and the norm $\|\mathbf{x}_i\|$ for every $v_i \in V_\mathrm{graph}$;        
    \item the claimed inner product $IP_i = \langle \mathbf{x}_i, \mathbf{x}_q\rangle$ and corresponding IPA proof $\rho_i$ for every $v_i \in V_\mathrm{graph}$;
    \item the corresponding Merkle proof for $V_\mathrm{graph}$;
    \item the DO's signature $\sigma$.

\end{itemize}

Upon receiving $(R, VO)$, the client first reconstructs the root hash and verifies the DO's signature $\sigma$. This step authenticates the search subgraph, including $ID_i$, $Adj_i$, $C_i$, and $\|\mathbf{x}_i\|$ for each visited node $v_i \in V_\mathrm{graph}$.
Subsequently, the client verifies each raw vector $\mathbf{x}_R$ in $R$ against its corresponding commitment $C_R$.
For each visited node $v_i \in V_\mathrm{graph}$, the client validates the IPA proof $\rho_i$ for the claimed inner product $IP_i$ by invoking $\funcname{VerifyIPA}(C_i, C_q, IP_i, \rho_i, pp)$. A valid IPA proof certifies that $IP_i = \langle \mathbf{x}_i, \mathbf{x}_q \rangle$ matches the commitment $C_i$.
Using the authenticated $IP_i$, the validated norm $\|\mathbf{x}_i\|$, and the locally computed $\|\mathbf{x}_q\|$, the client computes the distance $\delta(\mathbf{x}_i, \mathbf{x}_q)$ for each node $v_i$.
It then constructs a local search subgraph $\mathcal{I}'=\{(ID_i, Adj_i, \delta_i)\}_{v_i \in V_\mathrm{graph}}$ and replays the graph-based ANN search algorithm over $\mathcal{I}'$.
Unlike the strawman solution, which recomputes distances from high-dimensional raw vectors, \schemename replays the search using these authenticated distances. Verification succeeds if the replay yields the same result set $R=R'$ without attempting to access any node outside the authenticated set $V_\mathrm{graph}$.

\textit{\textbf{Example.}
The search in \Cref{fig-graph_with_search} visits $V_\mathrm{graph}=\{v_1,v_3,$\\$v_5,v_6,v_9,v_{11},v_{12}\}$ and returns $R=\{\mathbf{x}_9\}$. The SP includes $(ID_i,Adj_i,C_i,\|\mathbf{x}_i\|)$ for every $v_i\in V_\mathrm{graph}$ in the VO and provides the Merkle proof as in \Cref{sec:straw_search}. For distance authentication, the SP computes $IP_1=\langle\mathbf{x}_1,\mathbf{x}_q\rangle$ for $v_1$, generates the IPA proof $\rho_1$, and adds both to the VO instead of the raw vector $\mathbf{x}_1$.}
\textit{The client reconstructs the root from the node data and Merkle proof, verifies its signature, and checks each result vector in $R$ against its commitment (\eg, $\funcname{Com}(\mathbf{x}_9)=C_9$). For every node in $V_\mathrm{graph}$, it then verifies the IPA proof and recovers the distance. For $v_1$, it checks $\rho_1$ using $C_1$, $IP_1$, and the local query vector $\mathbf{x}_q$, then computes the Euclidean distance $\delta_1=(\|\mathbf{x}_1\|^2+\|\mathbf{x}_q\|^2-2IP_1)^{1/2}$.}

\subsection{Proof Aggregation}\label{sec-aggregation}
Although the IPA-based approach significantly reduces the VO size, it requires a distinct proof for every visited node $v_i \in V_\mathrm{graph}$. This imposes substantial computational overhead on the SP for proof generation and on the client for verification. Furthermore, high-recall ANN searches in machine learning workloads typically traverse thousands of nodes. The cumulative size and verification time of these individual proofs become a severe bottleneck. To address this, we propose an IPA aggregation protocol that compresses multiple proofs into a single proof. Specifically, it extends the original IPA with the following two PPT algorithms:

\textbullet~\funcname{ProveAggIPA}$(\mathbf{x}_q,\{\mathbf{x}_i\}_{i=1}^{n}, pp) \to (\{IP_i\}_{i=1}^n, \rho_{agg})$: Given a common query vector $\mathbf{x}_q$, a set of $n$ distinct vectors $\{\mathbf{x}_i\}_{i=1}^{n}$, and the public key $pp$, this algorithm computes the $n$ inner products $\{ IP_i = \langle \mathbf{x}_q, \mathbf{x}_i \rangle \}_{i=1}^n$ and generates a single aggregated IPA proof $\rho_{agg}$ that authenticates all of them simultaneously.

Our solution exploits the fact that all $n$ inner products share the same query vector $\mathbf{x}_q$ as a common input. Thus, we apply random linear combination~\cite{10.1007/978-3-642-38980-1_32} to batch these $n$ inner product arguments into one aggregated proof. Specifically, the prover first computes a random challenge $r$ using the Fiat-Shamir heuristic~\cite{fs} over the public transcript, which is derived from commitments and inner products corresponding to the query vector and $n$ input vectors:
\begin{align}
    r = \mathbb{H}(C_q \concat C_1 \concat IP_1 \concat\cdots \concat C_n \concat IP_n). \label{eq:r}
\end{align}
The prover then computes an aggregated witness vector:
\begin{align*}
    \mathbf{x}_{agg} = \sum_{i=1}^{n} r^i \cdot \mathbf{x}_i. 
\end{align*}
{Finally, the prover invokes the standard IPA proving routine:}
\begin{align*}
    (IP_{agg}, \rho_{agg}) \leftarrow \funcname{ProveIPA}(\mathbf{x}_{agg}, \mathbf{x}_q, pp),
\end{align*}
where $\rho_{agg}$ serves as the aggregated IPA proof.

\textbullet~\funcname{VerifyAggIPA}$(C_q,\{(C_i,IP_i)\}_{i=1}^{n},\rho_{agg},pp) \to \{0,1\}$: Given the commitment $C_q$ corresponding to the query vector $\mathbf{x}_q$, a set of $n$ pairs of commitments and inner products $\{ (C_i, IP_i) \}_{i=1}^n$, the aggregated IPA proof $\rho_{agg}$, and the public key $pp$, this algorithm outputs 1 if and only if $\langle \mathbf{x}_q, \mathbf{x}_i \rangle = IP_i \land \funcname{Com}(\mathbf{x}_i) = C_i$ for all $i \in \{1,\dotsc,n\}$.

The verification algorithm first recomputes the random challenge $r$ using \cref{eq:r}. It then derives the aggregated commitment $C_{agg}$ and the corresponding aggregated inner product $IP_{agg}$:
\begin{align*}
    C_{agg} = \sum_{i=1}^{n} r^i \cdot C_i, \quad
    IP_{agg} = \sum_{i=1}^{n} r^i \cdot IP_i.
\end{align*}
Finally, the verifier invokes the standard IPA verification routine:
\begin{align*}
    \funcname{VerifyIPA}(C_q, C_{agg}, IP_{agg}, \rho_{agg}, pp).
\end{align*}
The batch verification succeeds if and only if $\funcname{VerifyIPA}$ returns 1.

We now describe the integration of the aggregation protocol into the verifiable search procedure of \schemename. Compared to the previous construction, where the SP generates an individual IPA proof for every visited node, we modify the distance-authentication proof generation step.
Specifically, once the ANN search identifies the visited node set $V_\mathrm{graph}$, the SP no longer invokes \funcname{ProveIPA} independently for each $v_i \in V_\mathrm{graph}$. Instead, it aggregates the commitments $\{C_i\}_{v_i\in V_\mathrm{graph}}$ and the corresponding inner products $\{IP_i\}_{v_i\in V_\mathrm{graph}}$ and calls \funcname{ProveAggIPA} once, producing a single aggregated proof $\rho_{agg}$. Consequently, the multiple individual IPA proofs in the original VO are replaced by one compact proof, while the rest of the search algorithm remains unchanged.

During verification, instead of validating individual IPA proofs sequentially, the client invokes \funcname{VerifyAggIPA} to simultaneously authenticate all received inner products
 $\{IP_i\}_{v_i\in V_\mathrm{graph}}$ using $\rho_{agg}$. If this batch verification succeeds, the client accepts the integrity of all inner-product claims and proceeds with the remainder of the verification procedure.

\textit{\textbf{Example.} Continuing the previous example with $V_\mathrm{graph}=\{v_1,v_3,v_5,$ $v_6,v_9,v_{11},v_{12}\}$, rather than generating seven individual proofs $\{\rho_{1},\rho_{3},$ $\dotsc,\rho_{12}\}$, the SP runs \funcname{ProveAggIPA} once to produce a single proof $\rho_{agg}$ for the VO. During verification, the client computes $C_{agg}= rC_1 + r^2C_3 + \dotsm + r^7C_{12}$ and $IP_{agg} = rIP_1 + r^2IP_3 + \dotsm + r^7IP_{12}$, then uses $\rho_{agg}$ to simultaneously validate all inner-product claims before continuing with the rest of the verification.}

\section{Topology-aware \schemename}\label{sec:MeST}
This section proposes a novel ADS called the \underline{Me}rkle \underline{S}panning \underline{T}ree (\indname) for proximity graphs, which overcomes the topology-agnostic limitation mentioned in \Cref{sec:limitation} and seamlessly integrates with \schemename.

\subsection{Topology-aware ADS Construction}\label{sec:Mest_ADS}
Recall that the strawman solution constructs the MHT based on node identifiers, ignoring the topological structure of the proximity graph. Consequently, it fails to exploit the search index's topological semantics, leading to redundant and inefficient Merkle proofs in the VO. To address this, we propose building the MHT directly over the proximity graph to align the ADS with the search access patterns. However, optimizing VO size for ANN search poses two primary challenges:
\begin{itemize}
    \item \emph{Graph-to-tree mapping and path compression:} Unlike the hierarchical structure of an MHT, proximity graphs are undirected and often cyclic. The challenge lies in finding an optimal graph-to-tree mapping such that nodes frequently co-visited during search share a common Merkle path with high probability. This alignment is critical for maximizing proof compression by deduplicating shared internal hash nodes.
    \item \emph{Node content and minimal disclosure:} Since ANN search involves the iterative exploration of a node's neighborhood, the ADS must be carefully designed. We must determine the optimal granularity of information (\eg, adjacency lists and commitments) stored in each MHT node to ensure the client receives all necessary data for verification while minimizing the number of revealed Merkle nodes to reduce VO size.
\end{itemize}

We observe that graph-based ANN search typically employs greedy traversal guided by a priority queue, where exploring a node frequently leads to the subsequent exploration of its immediate neighbors. To capture this spatial-topological correlation, we construct the MHT based on a spanning tree rooted at the graph's global entry point.
While a standard unweighted minimum spanning tree can minimize hop distance by placing nodes reachable in $d$ hops at depth $d$ of the Merkle tree, it treats all neighbors equally and ignores the distance-driven nature of greedy search. Conversely, a traditional distance-weighted minimum spanning tree minimizes global edge weights without regard to tree depth, which may produce excessively deep paths and degrade Merkle proof efficiency.
To balance these requirements, we propose a Distance-aware Minimum Spanning Tree (\cref{alg:obfs}). This approach treats minimizing hop distance as the primary global objective while employing a local greedy strategy to prioritize the closest neighbors as a tie-breaker when multiple choices exist (Line~\ref{alg:obfs:chose}). By prioritizing local distance within a hop-constrained structure, the resulting ADS ensures that frequently co-visited nodes share common ancestor paths in the MHT, thereby maximizing proof compression in the VO.

\begin{algorithm}[t]
    \footnotesize
    \caption{Distance-aware Minimum Spanning Tree}
    \label{alg:obfs}
    \SetKwFunction{FBuildTree}{BuildSpanningTree}
    \SetKwProg{Fn}{Function}{:}{}
    \DontPrintSemicolon

    \Fn{\FBuildTree{$\mathcal{I}$}}{
        $Visited \leftarrow \{v_{\mathrm{ep}}\}$; $Cur \leftarrow \{v_{\mathrm{ep}}\}$;

        \While{$Cur \neq \varnothing$}{
            $Parent \leftarrow \varnothing$;\;
            \ForEach{$v_i \in Cur$, $v_j \in Adj_i$}{
                \lIf{$v_j \notin Visited$}{
                    $Parent[v_j] \leftarrow Parent[v_j] \cup \{v_i\}$;
                }
            }
            $Next \leftarrow \varnothing$;\;
            \ForEach{$v_j \in Parent$}{
                $v_i^* \gets\operatorname*{arg\,min}_{v_i \in Parent[v_j] } \delta(\mathbf{x}_i, \mathbf{x}_j)$;\label{alg:obfs:chose}\;
                Add edge $(v_i^*, v_j)$ to the spanning tree;\;
                $Visited \gets Visited \cup \{v_j\}$;
                $Next \gets Next \cup \{ v_j\}$;\;
            }
            $Cur \gets Next$;\;
        }
    }
\end{algorithm}

\begin{figure}[t]
    \centering
    \includegraphics[width=0.89\linewidth]{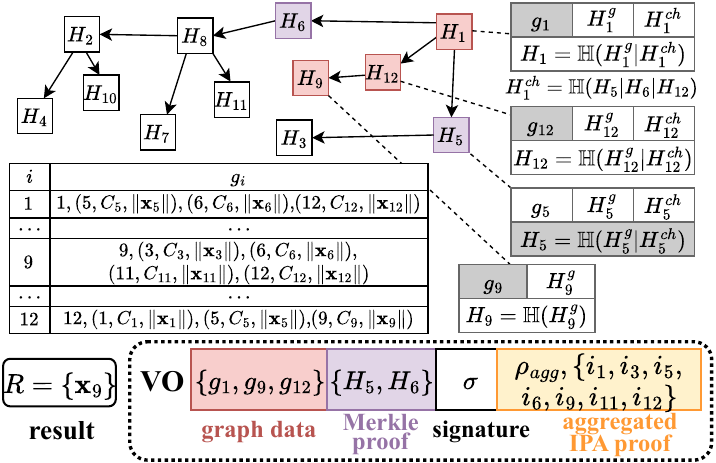}
    \caption{\indname Index}
    \label{fig-mesh_index}
\end{figure}

Having defined the spanning tree structure, we now address the second challenge: determining the optimal data to store in each Merkle tree node to support verifiable search while revealing minimal information. Because the Merkle spanning tree allows graph nodes to reside in both internal and leaf positions, we first define a standardized Graph Node ADS:
\begin{definition}[Graph Node ADS]
For each graph node $v_i$ in the ANN search index $\mathcal{I}$, its corresponding ADS $g_i$ consists of 
(i) $ID_i$, the identifier of the current node $v_i$;
(ii) $\{ (ID_j , C_j , \|\mathbf{x}_j\|) \}_{\forall v_j \in Adj_i}$, the identifiers, commitments, and norms of all immediate neighbors of $v_i$ in $\mathcal{I}$. %
The hash of $g_i$ is defined as
$
H_i^{g} = \mathbb{H}(ID_i \concat \{ ID_j \concat C_j \concat \|\mathbf{x}_j\| \}_{\forall v_j \in Adj_i}).$
\end{definition}

This design incorporates three key insights. First, the ADS $g_i$ is based on the original proximity graph $\mathcal{I}$ rather than the Merkle spanning tree, so ANN search traverses the original index $\mathcal{I}$ while the Merkle tree only authenticates the underlying data. Second, embedding each neighbor's commitment $C_j$ and norm $\|\mathbf{x}_j\|$ directly in $g_i$ lets the client compute and authenticate $\delta(\mathbf{x}_q,\mathbf{x}_j)$ immediately upon receiving $g_i$. The full neighbor ADS $g_j$ need not be revealed until that neighbor is visited, significantly reducing VO size. Finally, $H_i^g$ uses a flattened representation of all neighbors. Providing this information together matches greedy ANN search, which examines all neighbors of an explored node to update the priority queue, and avoids the overhead of nested hash functions.

The construction of the remaining ADS is as follows. For each leaf node in the Merkle spanning tree, the hash is defined as
$H_i = H_i^g$, directly using the corresponding graph node's ADS. In contrast, each internal node must encapsulate both its own graph node ADS and its children. Its hash is therefore defined as $H_i = \mathbb{H}(H_i^g \concat H_i^{ch})$, where $H_i^{ch}$ denotes the hash of the children. Because the Merkle spanning tree often exhibits varying and occasionally high fanout at internal nodes, we compute $H_i^{ch}$ using a threshold-based heuristic to optimize performance. If the fanout is below a predefined threshold, $H_i^{ch}$ is computed as a flattened list hash: $H_i^{ch} = \mathbb{H}(H_{i,ch_1} \concat \dotsm \concat H_{i,ch_m})$. Conversely, if the number of children is large, we construct a secondary binary MHT over these children and assign its root hash as $H_i^{ch}$. This hybrid approach ensures that when only a subset of children is visited during query processing, the verification remains efficient by minimizing the data required to reconstruct the Merkle root in the VO. \blue{Finally, to bind the entry point's own distance payload to the ADS, the DO computes $H_\mathrm{root}=\mathbb{H}(H_\mathrm{ep}\concat C_\mathrm{ep}\concat\|\mathbf{x}_\mathrm{ep}\|)$ and signs $H_\mathrm{root}$ as the trusted root of the entire ADS.}

\textit{\textbf{Example}. 
Following \Cref{fig-graph}, we first construct the spanning tree via \Cref{alg:obfs} starting from $v_1$.
The traversal first reaches nodes $\{v_5, v_6, v_{12}\}$ through $v_1$'s adjacency list. It next discovers $v_9$, reachable from both $v_6$ and $v_{12}$, and selects $v_{12}$ as its parent in the spanning tree by comparing $\delta(\mathbf{x}_9,\mathbf{x}_6)$ and $\delta(\mathbf{x}_9,\mathbf{x}_{12})$.
By applying this rule to all nodes, we obtain the final spanning tree structure shown in \Cref{fig-mesh_index}.
On top of this tree, each node stores the ADS of the corresponding graph node. 
Concretely, for node $v_1$, its graph neighbors are $v_5$, $v_6$, and $v_{12}$. Hence, its graph ADS is $g_1=\{1,(5,C_5,\|\mathbf{x}_5\|),(6,C_6,\|\mathbf{x}_6\|),(12,C_{12},\|\mathbf{x}_{12}\|)\}$, and the corresponding graph digest is $H_1^{g}=\mathbb{H}(g_1)$.}

\textit{At the tree level, a leaf node directly uses its graph digest as its node hash. For example, $v_9$ is a leaf in the spanning tree, so $H_9=H_9^{g}$. For a non-leaf node, \indname combines the graph-level digest with a digest of its children in the spanning tree. For example, the hash of node $v_1$ is computed as
$H_1=\mathbb{H}(H_1^{g}\concat H_1^{ch})$, where $H_1^{ch}=\mathbb{H}(H_5 \concat H_6 \concat H_{12})$ summarizes the hashes of the children of $v_1$ in the spanning tree. Applying this construction bottom-up to all nodes yields the root digest of the entire \indname index, namely $H_\mathrm{root}=\mathbb{H}(H_1\concat C_1\concat\|\mathbf{x}_1\|)$, which is then signed by the DO to generate $\sigma$.}

\subsection{Verifiable Search} \label{sec:MeST-verifiablesearch}
Next, we describe the verifiable search algorithm when integrated with MeST. As in \cref{sec:VIDA-search}, the SP executes a priority-queue-guided graph traversal. However, in addition to recording $V_\mathrm{graph}$ for all visited graph nodes, the SP also tracks a separate set $V_\mathrm{merkle}$ that contains the nodes visited in the Merkle tree. As detailed in \Cref{alg:mest_search}, $V_\mathrm{graph}$ captures all nodes when they are pushed into the priority queue and have their distance to the query evaluated (Line~\ref{alg:mest_search:graph}) to ensure correct search results. In contrast, $V_\mathrm{merkle}$ only tracks the subset of nodes that are actually \emph{explored} from the priority queue to examine their neighbors (Line~\ref{alg:mest_search:merkle}). By maintaining this distinction, the SP can generate an aggregated IPA proof for all data vectors in the larger $V_\mathrm{graph}$, while only requiring Merkle authentication paths for the significantly smaller $V_\mathrm{merkle}$, thereby minimizing the overall VO size. The SP then outputs each graph ADS $g_i$ for nodes in $V_\mathrm{merkle}$, a Merkle proof for $V_\mathrm{merkle}$, the original DO signature $\sigma$, all relevant inner products from $V_\mathrm{graph}$, and the aggregated IPA proof $\rho_{agg}$ as the VO.

On the client side, the verification procedure follows a logic similar to that in \cref{sec:VIDA-search}. The primary distinction lies in the handling of the two node sets, $V_\mathrm{merkle}$ and $V_\mathrm{graph}$.  The client first reconstructs the root hash using the Merkle proof to authenticate the received graph ADS ${\{g_i\}}_{v_i \in V_\mathrm{merkle}}$.  Since each graph ADS $g_i$ contains the commitments and norms for all its neighbors, the client can utilize them together with the inner products verified via $\rho_{agg}$ to reconstruct the search subgraph for every node in $V_\mathrm{graph}$. Finally, the client replays the ANN search over the locally reconstructed subgraph to verify the query result.

\begin{algorithm}[t]
    \footnotesize
    \caption{\indname Verifiable Search}
    \label{alg:mest_search}
    \DontPrintSemicolon
    \SetKwProg{Fn}{Function}{:}{}
    \SetKwFunction{FSearch}{Search}
    \SetKwFunction{FProof}{ProveMerkle}
    \SetKwFunction{FProveAgg}{ProveAggIPA}
    
    \Fn{\FSearch{$Q=(k,\mathbf{x}_q),\mathcal{I},\mathcal{A},\sigma$}}{
        $PriorityQueue \leftarrow \{(v_{\mathrm{ep}}, \delta_{\mathrm{ep}})\}$;
        $V_\mathrm{merkle} \gets \{v_\mathrm{ep}\}$;
        $V_\mathrm{graph}\gets\{v_\mathrm{ep}\}$;\;
        \While{$(v_i,\delta_i)\leftarrow PriorityQueue.\mathrm{dequeue}()$}{
            Update result set $R$ with $v_i$ and stop if there are already $k$ closer results;\;
            $V_{\mathrm{merkle}} \leftarrow V_{\mathrm{merkle}} \cup \{v_i\}$;\label{alg:mest_search:merkle}\;
            \ForEach{$v_j \in Adj_i$}{
                $V_\mathrm{graph} \leftarrow V_\mathrm{graph} \cup \{v_j\}$;\label{alg:mest_search:graph}\;
                $PriorityQueue.\mathrm{enqueue}((v_j,\delta_j))$;\label{alg:mest_search:priorityqueue}\;
            }
        }
        
        $\pi \leftarrow$ \FProof{$V_{\mathrm{merkle}},\mathcal{A}$};\;
        $\rho_{agg} \leftarrow$ \FProveAgg{$\mathbf{x}_q,\{\mathbf{x}_i\}_{v_i \in V_\mathrm{graph}}$};\;
        $VO \leftarrow (\{g_i\}_{v_i \in V_{\mathrm{merkle}}},\pi, \sigma,\{IP_i\}_{v_i \in V_\mathrm{graph}},\rho_{agg})$;\; 
        \KwRet $(R,VO)$.\;
    }
\end{algorithm}

\textit{\textbf{Example.}
Consider the search process in \Cref{fig-graph_with_search}. 
The search visits $V_\mathrm{graph}=\{v_1,v_3,v_5,v_6,v_9,v_{11},v_{12}\},$
while only the proven set $V_\mathrm{merkle}=\{v_1,v_9,v_{12}\}$ needs to be covered by the Merkle proof in \indname. As shown by the graph ADS table in \Cref{fig-mesh_index}, the remaining visited nodes $V_\mathrm{graph}\setminus V_\mathrm{merkle}=\{v_3,v_5,v_6,v_{11}\}$ are all exposed through the authenticated graph ADS of the proven nodes $g_1$, $g_9$, and $g_{12}$. Accordingly, the VO contains the graph ADS $\{g_1,g_9,g_{12}\}$, the required sibling hashes $\{H_5,H_6\}$, the signature $\sigma$, the claimed inner products $\{IP_i\}_{v_i\in V_\mathrm{graph}}$, and the aggregated IPA proof $\rho_{agg}$.}

\textit{During verification, the client first reconstructs the Merkle root from $\{g_1,g_9,g_{12}\}$ and $\{H_5,H_6\}$, then verifies the signature $\sigma$. It next verifies $\rho_{agg}$ and recovers the distances of all nodes in $V_\mathrm{graph}$. Using the authenticated graph ADS of $V_\mathrm{merkle}$ together with the recovered distances for $V_\mathrm{graph}$, the client constructs the local subgraph $\mathcal{I}'$ and replays the search.}

\subsection{\indname for Multi-layered Graphs}

\blue{Many graph-based ANN indexes, such as HNSW~\cite{HNSW}, use multi-layered graphs. \indname supports them by building the Merkle spanning tree over the union of all layer-wise edges and organizing each node's graph ADS into layer-specific neighbor payloads whose digests are combined into the node hash. During verifiable search, the SP reveals only accessed-layer payloads and provides digests for the others. The client replays the original top-down search and rejects if any required payload is missing. %
Full details are in \iftechreport\Cref{append-mest-multilayer}\else our technical report~\cite{fullversion}\fi.}

\section{Quantization-aware \schemename}\label{sec:qua}

\blue{\schemename and \indname reduce authentication overhead through IPA aggregation and a topology-aware ADS, respectively. Nevertheless, each evaluated candidate introduces an exact-distance authentication instance for verifier-side replay, so the remaining cost scales with the candidate count. Large-scale and high-recall graph-based ANN searches may evaluate many candidates, shifting the dominant authentication cost from per-candidate overhead to the number of such instances.}

\blue{Modern ANN systems address an analogous query-time bottleneck through vector quantization, which encodes vectors into compact representations and evaluates candidates without accessing their full-precision vectors. Classical methods like PQ~\cite{pq} and SQ~\cite{sq} approximate distances from compressed codes and often rerank a shortlist using exact distances to preserve search accuracy. In contrast, methods such as RaBitQ~\cite{rabitq} compute one-sided lower bounds that enable pruning when they exceed the current search threshold. Such pruning is more compatible with authenticated search, because the verifier can recompute the bound and validate the pruning decision without checking the exact distance.}

\blue{Motivated by this observation, we extend \schemename to authenticate quantization-aware graph search. Exact distances are authenticated only for candidates that require exact evaluation, whereas candidates eliminated by quantization can be justified using compact authenticated quantization payloads. Quantization thus reduces both ANN query-time computation and the number of exact distance computations that must enter the aggregated IPA proof. We next present the corresponding ADS construction and verifiable search protocol.}

\blue{\textbf{ADS Construction.}
To support quantization-aware authentication, we extend the Graph Node ADS defined in \Cref{sec:Mest_ADS}. Besides the exact-distance payload used for IPA-based authentication, each neighbor entry must also bind the quantization payload used to compute the lower bound. 
Accordingly, each neighbor $v_j$ contributes two payloads to the quantization-aware Graph Node ADS: the exact-distance payload $(C_j,\|\mathbf{x}_j\|)$ and the quantization payload $\eta_j$.} \blue{All other components of the ADS and its construction are unchanged from \Cref{sec:Mest_ADS}.}

\begin{definition}[Quantization-aware Graph Node ADS]
\blue{For each graph node $v_i$ in the index $\mathcal{I}$, the ADS $g_i$ contains:
(i) $ID_i$, the identifier of the current node $v_i$;
(ii) $\{ (ID_j , C_j , \|\mathbf{x}_j\|,\eta_j )\}_{\forall v_j \in Adj_i}$, the identifiers, commitments, norms, and quantization payloads of all immediate neighbors of $v_i$ in $\mathcal{I}$.
The hash of $g_i$ is defined as
$
H_i^{g} = \mathbb{H}(ID_i \concat \{ ID_j \concat H_j^{exa} \concat H_j^{qua}  \}_{\forall v_j \in Adj_i})
$, where
$
H_j^{exa} = \mathbb{H}(C_j \concat \|\mathbf{x}_j\|)
$
is the hash of the exact-distance payload of $v_j$, and 
$
H_j^{qua} = \mathbb{H}(\eta_j)
$
is the hash for the quantization payload.}
\end{definition}

\begin{algorithm}[t]
    \footnotesize
    \caption{Quantization-aware Verifiable Search}
    \label{alg:qann_search}
    \DontPrintSemicolon
    \SetKwProg{Fn}{Function}{:}{}
    \SetKwFunction{FRSearch}{QuantAwareSearch}
    \SetKwFunction{FLB}{QuantLB}
    \SetKwFunction{FDist}{ExactDist}
    \SetKwFunction{FProveAgg}{ProveAggIPA}
    
    \Fn{\FRSearch{$Q=(k,\mathbf{x}_q),\mathcal{I},\mathcal{A},\sigma$}}{       
        $PriorityQueue \leftarrow \{(v_{\mathrm{ep}},\delta_{\mathrm{ep}})\}$;\;
        $V_{\mathrm{merkle}} \leftarrow \{v_{\mathrm{ep}}\}$;
        $V_{\mathrm{exa}} \leftarrow \{v_{\mathrm{ep}}\}$;
        $V_{\mathrm{qua}} \leftarrow \varnothing$;\;
        
        \While{$(v_i,\delta_i)\leftarrow PriorityQueue.\mathrm{dequeue}()$}{
            Update result set $R$ with $v_i$ and stop if there are already $k$ closer results;\;
            $V_{\mathrm{merkle}} \leftarrow V_{\mathrm{merkle}} \cup \{v_i\}$;\label{alg:qann_search:merkle}\;
            \ForEach{$v_j \in Adj_i$}{
                $\delta_{\mathrm{th}} \leftarrow PriorityQueue.\mathrm{threshold}()$;\label{alg:qann_search:threshold}
                $LB_j \leftarrow$ \FLB{$\eta_j,\mathbf{x}_q$};\label{alg:qann_search:lb}
                
                \If{$LB_j > \delta_{\mathrm{th}}$}{
                    $V_{\mathrm{qua}} \leftarrow V_{\mathrm{qua}} \cup \{v_j\}$;\label{alg:qann_search:v_qua}
                    \textbf{continue};\;
                }
                
                $\delta_j \leftarrow$ \FDist{$\mathbf{x}_j,\mathbf{x}_q$};
                $V_{\mathrm{exa}} \leftarrow V_{\mathrm{exa}} \cup \{v_j\}$;\;\label{alg:qann_search:v_exa}
                $PriorityQueue.\mathrm{enqueue}((v_j,\delta_j))$;\label{alg:qann_search:enqueue}
                
            }
        }
    $\pi \leftarrow$ \FProof{$V_{\mathrm{merkle}},\mathcal{A}$};\;
        $\rho_{agg} \leftarrow$ \FProveAgg{$\mathbf{x}_q,\{\mathbf{x}_i\}_{v_i \in V_\mathrm{exa}}$};\;
        $VO \leftarrow (\{g_i^{{op}}\}_{v_i \in V_{\mathrm{merkle}}},\pi,\sigma,\{IP_i\}_{v_i \in V_{\mathrm{exa}}},\rho_{agg})$;\label{alg:qann_search:VO}\;
        \KwRet $(R,VO)$.\;
    }
\end{algorithm}

\begin{figure}[t]
    \centering
    \includegraphics[width=0.9\linewidth]{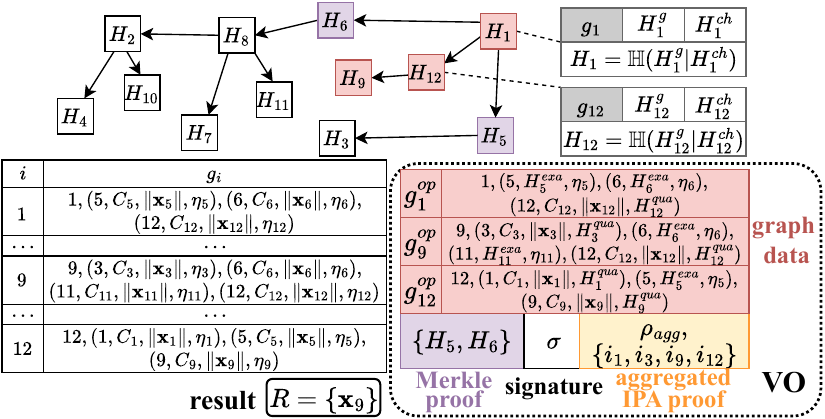}
    \caption{Quantization-aware \indname Index}
    \label{fig-mesh_index_qua}
\end{figure}

\blue{\textbf{Verifiable Search.}
During query processing, the SP executes the quantization-aware graph search in \Cref{alg:qann_search}. The search follows the same priority-queue-guided traversal as the original graph search. When a node $v_i$ is dequeued and expanded, the SP adds it to $V_{\mathrm{merkle}}$ and scans each neighbor $v_j\in Adj_i$. It first obtains the current search threshold $\delta_{\mathrm{th}}$ and computes the lower bound $LB_j$ from $\eta_j$ and $\mathbf{x}_q$ (Line~\ref{alg:qann_search:threshold}). If $LB_j>\delta_{\mathrm{th}}$, then $v_j$ is pruned without exact distance computation and recorded in the quantization-pruned node set $V_{\mathrm{qua}}$ (Line~\ref{alg:qann_search:v_qua}). Otherwise, the SP computes the exact distance $\delta_j$, adds $v_j$ to the distance-evaluated node set $V_{\mathrm{exa}}$, and inserts it into the priority queue (Lines~\ref{alg:qann_search:v_exa}--\ref{alg:qann_search:enqueue}). If a node is encountered multiple times, membership in $V_{\mathrm{exa}}$ takes precedence: once its exact distance is evaluated, the node is excluded from $V_{\mathrm{qua}}$.
Then, for each $v_i\in V_{\mathrm{merkle}}$, the SP materializes a node opening $g_i^{op}$ that discloses only the payload required for each exposed neighbor. Distance-evaluated neighbors reveal $(C_j,\|\mathbf{x}_j\|)$, while quantization-pruned neighbors reveal $\eta_j$. The unopened component is represented by its digest. %
The SP then generates an aggregated IPA proof for the nodes in $V_{\mathrm{exa}}$ and constructs the VO.} %

\blue{Client-side verification is similar to that described in \Cref{sec:MeST-verifiablesearch}, except that the distance-evaluated and quantization-pruned branches are handled separately. The client first reconstructs the graph node digests from $\{g_i^{op}\}_{v_i\in V_{\mathrm{merkle}}}$ and then reconstructs the root hash. For nodes in $V_{\mathrm{exa}}$, it verifies $\rho_{agg}$ and recovers their authenticated distances. For each
$v_j\in V_{\mathrm{qua}}$, it computes $LB_j$ from the opened $\eta_j$ and checks $LB_j>\delta_{\mathrm{th}}$ at the same replay point. Finally, the client replays the search to verify the ANN result.}

\blue{\textit{\textbf{Example}. Consider the search in \Cref{fig-graph_with_search}. Suppose
$V_\mathrm{qua}=\{v_5,v_6,v_{11}\}$ is the quantization-pruned set, while the remaining visited nodes $V_\mathrm{exa}=\{v_1,v_3,v_9,v_{12}\}$ are evaluated using exact distances. As shown in \Cref{fig-mesh_index_qua}, the node openings $g_1^{{op}}, g_9^{{op}}, g_{12}^{{op}}$ are included in the VO. Within each node opening, every exposed neighbor is disclosed according to the branch taken during the quantization-aware search. For example, $6,H_6^{{exa}},\eta_6$ and $9,C_9,\|\mathbf{x}_9\|,H_9^{{qua}}$ are exposed differently.}
\textit{During verification, the client reconstructs the root and verifies the signature $\sigma$ and proof $\rho_{{agg}}$. For each node in $V_{\mathrm{qua}}$, the client recomputes $LB_j$ from the opened quantization payload $\eta_j$. Finally, the client replays the search using exact distances for $V_{\mathrm{exa}}$ and lower bounds for $V_{\mathrm{qua}}$.}
}

\section{Security Analysis}\label{sec:analysis}
We first establish the computational knowledge soundness of IPA aggregation using the underlying IPA extractor~\cite{bulletproofs}, then use authenticated distances and deterministic replay to prove the execution integrity of \schemename and its variants. Complete proofs are provided in
\iftechreport
\Cref{append-soundness,append-security}.
\else
our technical report~\cite{fullversion}.
\fi

\begin{restatable}[Soundness of IPA Aggregation]{theorem}{AggSoundness}\label{thm:soundness}
Let $\textsf{Adv}$ be a PPT adversary that produces a valid aggregated IPA proof for the relation $\langle \mathbf{x}_{agg},\mathbf{x}_q \rangle = IP_{agg}$ with non-negligible probability. There exists a PPT extractor $\mathcal{E}$ that extracts the witnesses $\{\mathbf{x}_i\}_{i=1}^n$ such that $\langle \mathbf{x}_i, \mathbf{x}_q \rangle = IP_i$ for all $i \in \{1, \dotsc, n\}$.
\end{restatable}

\begin{proof}[Proof Sketch]
By rewinding the prover at the Fiat--Shamir challenge~\cite{forkinglemma}, $\mathcal{E}$ obtains accepting transcripts for $n$ distinct nonzero challenges. The IPA extractor yields an aggregated witness per transcript. Solving the resulting invertible Vandermonde system recovers the individual witnesses. Commitment homomorphism and binding tie each $\mathbf{x}_i$ to $C_i$, while inner-product linearity gives $\langle\mathbf{x}_i,\mathbf{x}_q\rangle=IP_i$.
\end{proof}

We next consider execution integrity under Definition~\ref{def:security}, including the topology-aware and quantization-aware variants.

\begin{restatable}{theorem}{VIDASecurity}
\label{thm:security}
\schemename and its variants are secure with respect to Definition~\ref{def:security}, assuming the underlying cryptographic primitives, including IPA aggregation, are secure.
\end{restatable}

\begin{proof}[Proof Sketch]
An accepted forgery must either alter the authenticated data or deviate from the designated search. Altering graph structure or metadata requires a hash collision or signature forgery; changing a committed vector violates binding; and falsifying an authenticated distance contradicts \Cref{thm:soundness}. Otherwise, replay uses genuine graph data and distances: omitting a required node causes rejection, while deterministic replay yields the honest result. \indname preserves this argument through authentication against the signed root. For the quantization-aware variant, pruning is checked using authenticated lower-bound payloads at the same replay threshold. Under the quantization guarantee, a valid pruning step, or its replacement by an exact evaluation, preserves the effective search state. Thus, an incorrect result can be accepted only with negligible probability.
\end{proof}

\section{Evaluation}\label{sec:evaluation}
\subsection{Experiment Settings}
\textbf{Datasets and Query Workload.} 
We conduct extensive experiments on six datasets with diverse sizes, dimensions, and data types, as summarized in Table~\ref{table-dataset}. The synthetic \textsf{Cluster} dataset is generated following~\cite{graph_benchmark_21}. %
Regarding query workloads, we use the official query sets when available. For datasets without a predefined query set, we randomly sample 1,000 vectors from the base data to serve as queries. Unless otherwise specified, all experiments use Euclidean distance as the similarity metric and set the result size to $k=100$. The target recall for each dataset follows the varying-$k$ experiment.%

\noindent
\textbf{Metrics.} 
In our experiments, we primarily focus on end-to-end latency, defined as the total time elapsed from the client sending a query until completion of the verification process.
We report recall to evaluate search accuracy. Furthermore, to assess index construction overhead and storage efficiency, we measure the construction time and the size of the ADS.

\begin{table}[t]
    \small
    \centering
    \caption{{Dataset Statistics}}
    \label{table-dataset}
    \resizebox{\columnwidth}{!}{
    \begin{tabular}{c|cccc}
        \hline
        \textbf{Dataset} & \textbf{Size} & \textbf{Dimension} & \textbf{Query Size} & \textbf{Type} \\
        \hline
        OpenAI3072~\cite{openai3072} & 999,000 & 3,072 & 1,000 & Text \\
        WIT~\cite{WIT} & 6,476,255 & 2,048 & 1,000 & Image \\
        LAION~\cite{LAION} & 200,000,000 & 512 & 1,000 & Image \\
        Wikipedia~\cite{Wikipedia} & 106,712,786 & 1,024 & 1,000 & Text \\
        Msmarco~\cite{Msmarco} & 113,520,750 & 1,024 & 1,677 & Text \\
        Cluster & 10,000,000 & 4,096 & 1,000 & Synthetic \\
        \hline
    \end{tabular}
    }
\end{table}

\begin{figure*}[t]
    \centering
    \makeatletter
    \renewcommand{\@IEEEfigurecaptionsepspace}{\vskip2pt\relax\nointerlineskip}
    \makeatother
    \begin{minipage}{\linewidth}
        \centering
        \includegraphics[width=0.975\linewidth]{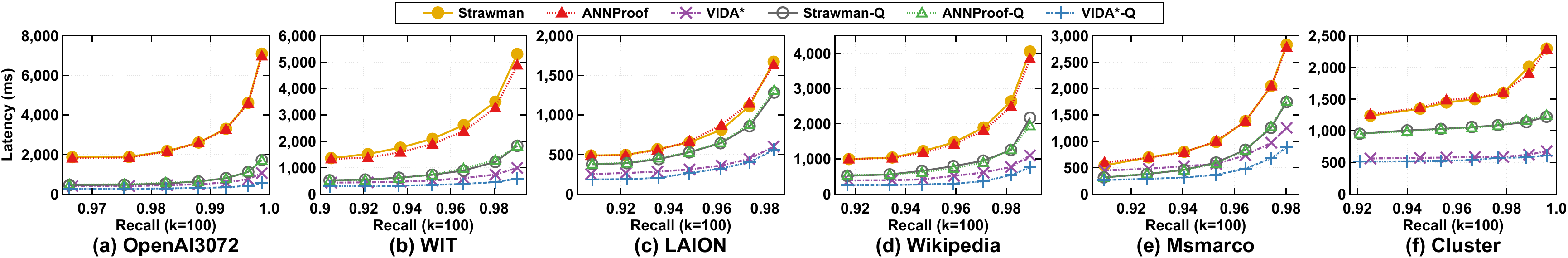}
        \caption{Latency on Different Recalls}
        \label{fig:latency_recall}
    \end{minipage}\par\vspace{2pt}
    \begin{minipage}{\linewidth}
        \centering
        \includegraphics[width=0.54\linewidth]{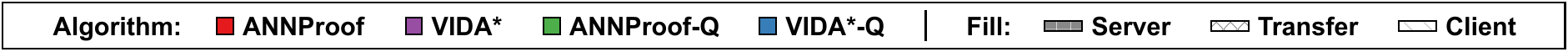}
    \end{minipage}\par\nointerlineskip\vspace{1pt}

    \begin{minipage}[b]{0.487\linewidth}
        \centering
        \includegraphics[width=\linewidth,height=0.3193\linewidth,trim=0 16.5bp 0 29bp,clip]{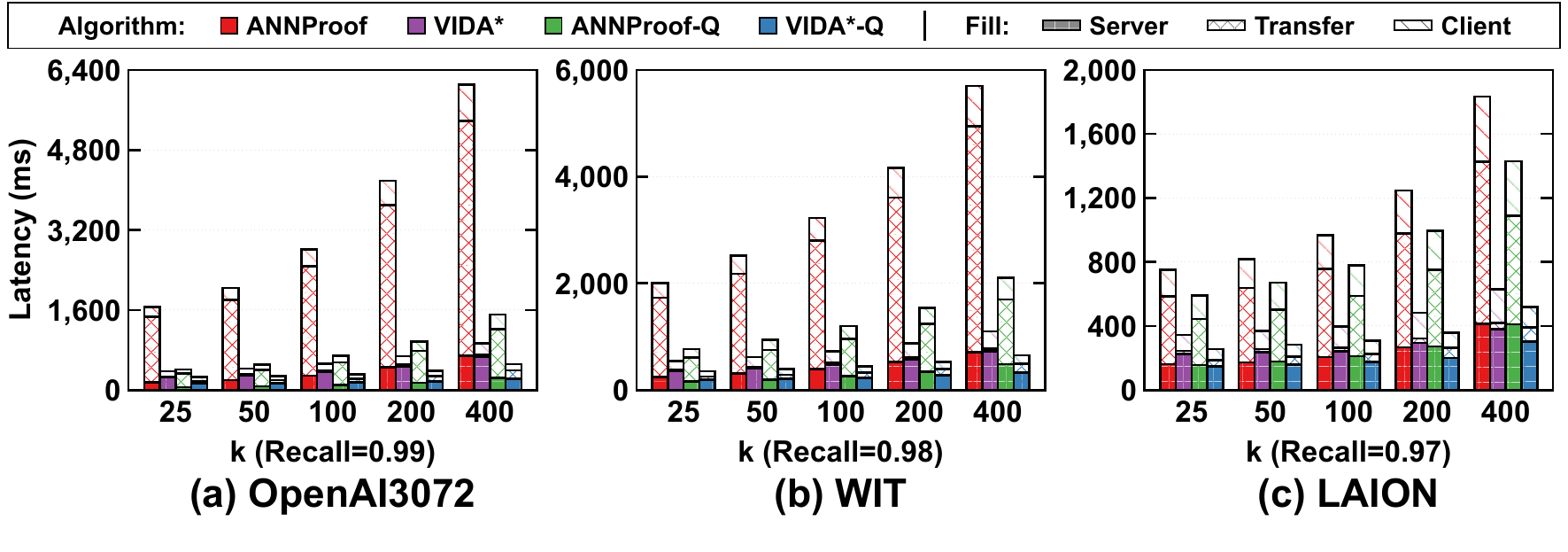}
        \caption{Latency of Varying $k$}
        \label{fig:latency_k}
    \end{minipage}
    \begin{minipage}[b]{0.487\linewidth}
        \centering
        \includegraphics[width=\linewidth,trim=0 0 0 29bp,clip]{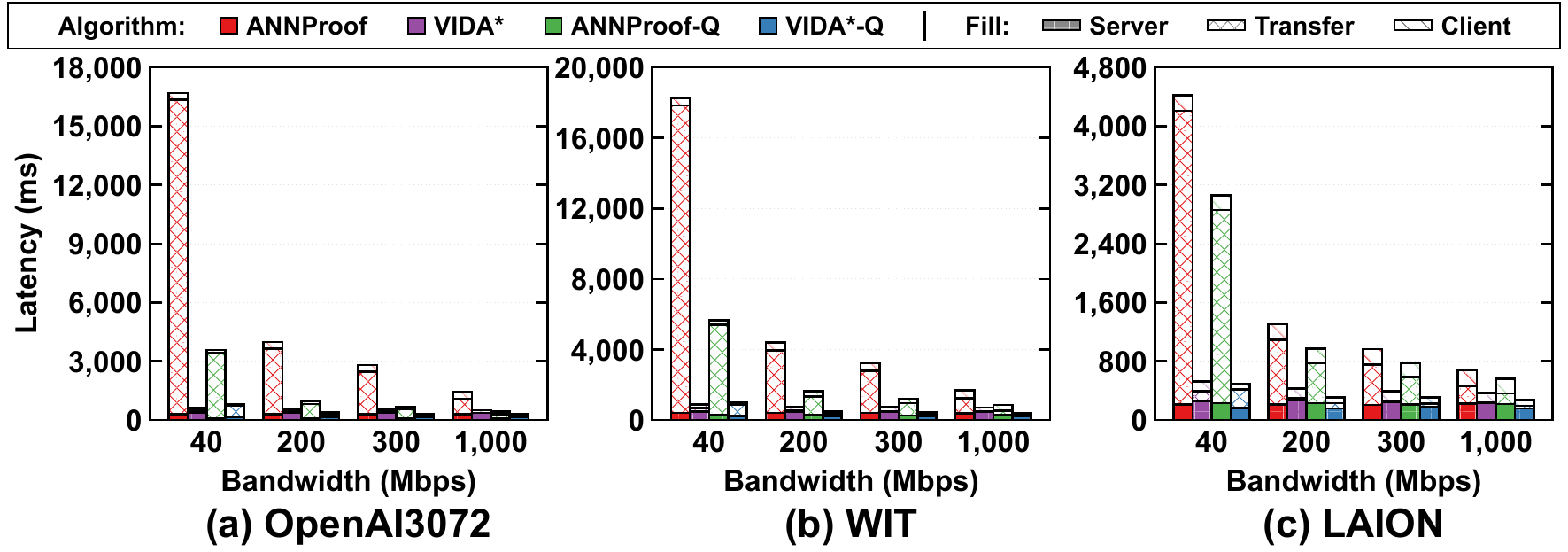}
        \caption{Latency on Different Network Settings}
        \label{fig:latency_bandwidth}
    \end{minipage}\par\vspace{2pt}
    
    \begin{minipage}{\linewidth}
        \centering
        \includegraphics[width=0.972\linewidth]{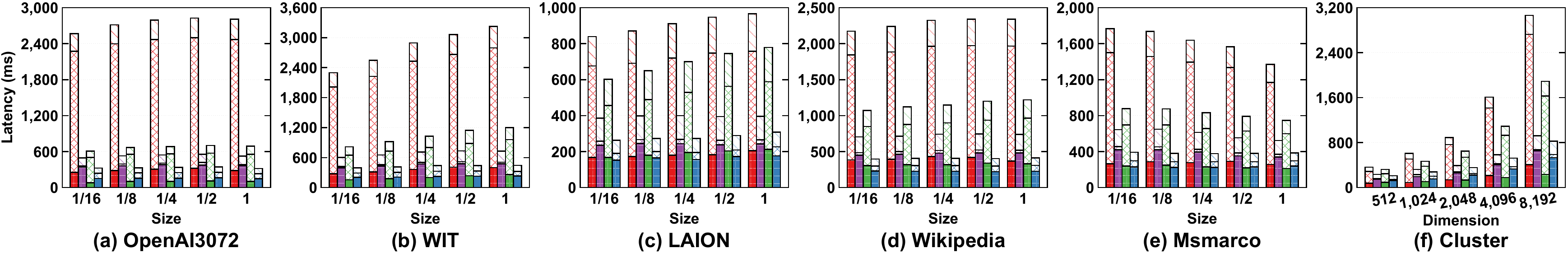}
        \caption{Scalability Test}
        \label{fig:latency_scal}
    \end{minipage}
\end{figure*}

\noindent
\blue{
\textbf{Main Algorithms for Evaluation.}
We evaluate three base methods and their quantization-aware variants.
\textbf{Strawman} employs a standard MHT over the graph and transmits raw vectors for distance verification.
\textbf{ANNProof}~\cite{annproof_baseline1} improves Strawman with a sharded MHT, while retaining raw-vector-based verification.
\textbf{VIDA*} is the full topology-aware instantiation of VIDA described in \Cref{sec:MeST}, combining aggregated IPA with MeST.
Their quantization-aware variants, \textbf{Strawman-Q}, \textbf{ANNProof-Q}, and \textbf{VIDA*-Q}, apply the quantization-aware search in \Cref{sec:qua}, using authenticated quantization payloads for pruned candidates and their respective original mechanisms for exact-distance verification.}

We use HNSW~\cite{HNSW} as the underlying ANN search index, unless otherwise specified. For the HNSW index configuration, we set the construction parameter $ef_{\text{construction}}$ to 500 across all datasets. The maximum number of links per node, $M$, is set to 32. %
During the query phase, we vary $ef_{\text{search}}$ to achieve different trade-offs between recall and latency. For IPA, we instantiate the protocol using the lightweight Bulletproofs~\cite{linearproof} (a variation) together with the Springproofs~\cite{springproofs} framework. \blue{For quantization-aware variants, we use RaBitQ~\cite{rabitq} on top of the corresponding graph index.}

\noindent
\textbf{Environment.} We implement the proposed schemes and baselines in the Rust programming language.
The source codes are available at \url{https://github.com/hkbudb/auth-ann}.
For the cryptographic primitives, the IPA is implemented via Bulletproofs~\cite{bulletproofs_library} using the Ristretto255 curve for group operations. 
All experiments were conducted on a server equipped with Dual 28-core Intel Xeon Gold 6330 @ 2.0GHz CPUs and 2TB RAM. To simulate real-world outsourced database scenarios, we limit the network bandwidth using the toxiproxy utility~\cite{toxiproxy}. The default network bandwidth is set to 300 Mbps.

\subsection{Search Performance}

\subsubsection{Latency under Varying Recall Targets}
We first evaluate the trade-off between end-to-end query latency and recall. \Cref{fig:latency_recall} compares the methods across recall targets on six datasets. For \textsf{Msmarco}, we use NSG~\cite{nsg} as the underlying ANN algorithm to evaluate our approach's generality beyond HNSW. We also include Strawman and its quantization-aware variant Strawman-Q as reference baselines. As shown in \Cref{fig:latency_recall}, Strawman(-Q) and ANNProof(-Q) show very similar latency trends. We therefore omit Strawman and use ANNProof as the baseline in subsequent experiments.

\blue{Overall, VIDA* consistently outperforms Strawman and ANNProof across all datasets, and the advantage becomes more pronounced in high-recall settings. This trend is especially evident on high-dimensional \textsf{OpenAI3072}. At the highest recall target, VIDA* achieves a 6.6$\times$ speedup over ANNProof, while VIDA*-Q achieves a 3.0$\times$ speedup over ANNProof-Q.
These results are consistent with our analysis. Achieving higher recall requires increasing the search parameter, \eg, $ef_{\text{search}}$ for HNSW, which increases the number of nodes visited during graph traversal. For raw-vector-based baselines, the transmission cost grows with the number of visited nodes and becomes the dominant contributor to end-to-end latency. In contrast, VIDA* replaces raw-vector transmission with compact commitments and aggregated IPA proofs, and further uses \indname to place neighboring graph nodes close together in the authentication tree. This reduces both distance-authentication cost and structural proof size. For instance, on \textsf{OpenAI3072}, VIDA* reduces the transfer cost by 99.2\% compared with ANNProof at the highest recall target, while VIDA*-Q reduces it by 82.5\% compared with ANNProof-Q.
Quantization-aware variants further reduce latency by reducing the number of candidates requiring exact-distance authentication. Compared with VIDA*, VIDA*-Q achieves additional speedups of 1.9$\times$, 1.7$\times$, and 1.4$\times$ on \textsf{OpenAI3072}, \textsf{WIT}, and \textsf{Wikipedia}, respectively, confirming that quantization complements our cryptographic and topology-aware optimizations.}

\subsubsection{Latency under Varying \texorpdfstring{$k$}{k}}
We vary $k$ from 25 to 400 and tune the search parameters to maintain each dataset's target recall. \Cref{fig:latency_k} decomposes latency on \textsf{OpenAI3072}, \textsf{WIT}, and \textsf{LAION} into server-side verifiable search (Server), network transfer (Transfer), and client-side verification (Client). Results for the other datasets are in \iftechreport\Cref{append:additional-evaluation}\else our technical report~\cite{fullversion}\fi.
\blue{
Across all six datasets, VIDA* and VIDA*-Q consistently outperform their corresponding baselines across all $k$ values, achieving up to 6.5$\times$ and 3.4$\times$ speedups over ANNProof and ANNProof-Q, respectively. The gains are especially pronounced on high-dimensional datasets. On \textsf{OpenAI3072} with $k=400$, ANNProof and ANNProof-Q incur 6,109 ms and 1,506 ms, respectively, whereas VIDA*-Q requires only 519 ms and reduces transfer cost by 82.5\% over ANNProof-Q. This confirms that our quantization-aware authentication further reduces communication overhead, even against the quantization-aware baseline.}

Latency generally increases with $k$ as more nodes are visited and larger VOs are generated, but the growth is nonlinear. At a fixed recall, small-k queries may still require large search parameters, so node visits do not decrease proportionally with $k$. This effect makes the raw-vector-based baselines costly even at small $k$, whereas our schemes grow more slowly. On \textsf{OpenAI3072}, when $k$ increases from 25 to 400, the latency of VIDA*-Q increases by only 2.0$\times$, compared with 2.5$\times$ for VIDA*, 3.7$\times$ for ANNProof, and 3.6$\times$ for ANNProof-Q. %
We also test smaller $k$ at lower recall targets on this dataset
(see \iftechreport\Cref{append:applicability}\else our technical report~\cite{fullversion}\fi).

\subsubsection{Latency under Varying Network Bandwidth}
We next evaluate performance under different network conditions by simulating four bandwidth profiles~\blue{\cite{yang2022mobile}}: 4G (40 Mbps), 5G (200 Mbps), Broadband (300 Mbps), and Server (1,000 Mbps). 
\blue{
\Cref{fig:latency_bandwidth} reports results on three datasets, while results for the other datasets are in \iftechreport\Cref{append:additional-evaluation}\else our technical report~\cite{fullversion}\fi. ANNProof and ANNProof-Q are highly sensitive to bandwidth degradation. Their latency increases sharply as available bandwidth decreases, indicating that large-scale data transmission remains the dominant bottleneck. For example, on \textsf{OpenAI3072}, when moving from Server bandwidth to 4G, the latency of ANNProof and ANNProof-Q increases by 11.7$\times$ and 8.5$\times$, respectively.}

\blue{In contrast, VIDA* and VIDA*-Q remain much more robust under bandwidth-limited settings. On \textsf{OpenAI3072}, the latency of VIDA* increases by only 1.2$\times$ when moving from 1,000 Mbps to 40 Mbps, while VIDA*-Q increases by 2.8$\times$. Under the most constrained 4G setting, ANNProof and ANNProof-Q incur 16.7~s and 3.6~s latency, respectively, whereas VIDA* and VIDA*-Q reduce the latency to 0.63~s and 0.82~s. Compared with ANNProof-Q, VIDA*-Q reduces the transfer cost by 82.8\% on \textsf{OpenAI3072} and by 87.3\% on \textsf{WIT}. These results show that compact distance authentication and topology-aware proofs substantially reduce the communication bottleneck, while quantization-aware authentication remains effective against its corresponding quantized baseline.
}

\subsection{Scalability and Efficiency}
We next evaluate the efficiency of the proposed schemes from three perspectives: scalability, offline construction overhead, and the contribution of \blue{the core optimization mechanisms}. Unless otherwise specified, we fix the search parameters across different methods.

\subsubsection{Scalability Analysis}\label{sec:scal}
We first study how query latency varies with dataset cardinality. To this end, we construct subsets of different sizes by randomly sampling from five real-world datasets. As shown in \Cref{fig:latency_scal}(a)-(e), for datasets such as \textsf{OpenAI3072} and \textsf{Wikipedia}, query latency increases only moderately as the dataset size grows, which is consistent with the expected logarithmic scaling behavior of graph-based indexes. An exception is the \textsf{Msmarco} dataset. For this sparse and high-dimensional dataset, random sampling degrades graph navigability, forcing the search procedure to visit more nodes to maintain the target recall.

\blue{We then evaluate the impact of vector dimensionality using the synthetic \textsf{Cluster} dataset. We fix the dataset size and vary the vector dimension $d$ from 512 to 8,192. \Cref{fig:latency_scal}(f) shows that ANNProof is highly sensitive to vector dimensionality. Its latency increases by 8.5$\times$, while ANNProof-Q still increases by 5.8$\times$. In contrast, VIDA* and VIDA*-Q increase by only 3.9$\times$ and 3.9$\times$, respectively. This result suggests that our IPA-based distance authentication significantly reduces the dependence of verification cost on vector dimensionality, making our schemes particularly well suited for modern high-dimensional embedding workloads.}
We also evaluate query latency at lower dimensions
(see \iftechreport\Cref{append:applicability}\else our technical report~\cite{fullversion}\fi).

\subsubsection{Offline Construction}
\blue{We evaluate offline index initialization in terms of construction time and ADS size. Relative to their corresponding baselines, VIDA* and VIDA*-Q require 5.5$\times$ and 5.4$\times$ the construction time on average, with ADS size increases of 116.1\% and 75.4\%, respectively. This one-time cost is paid by the data owner in exchange for the online latency and communication reductions. Detailed results and analysis are in \iftechreport\Cref{append:offline-construction}\else our technical report~\cite{fullversion}\fi.}

\begin{figure}[t]
    \centering
    \begin{minipage}{\linewidth}
        \centering
        \includegraphics[width=0.92\linewidth]{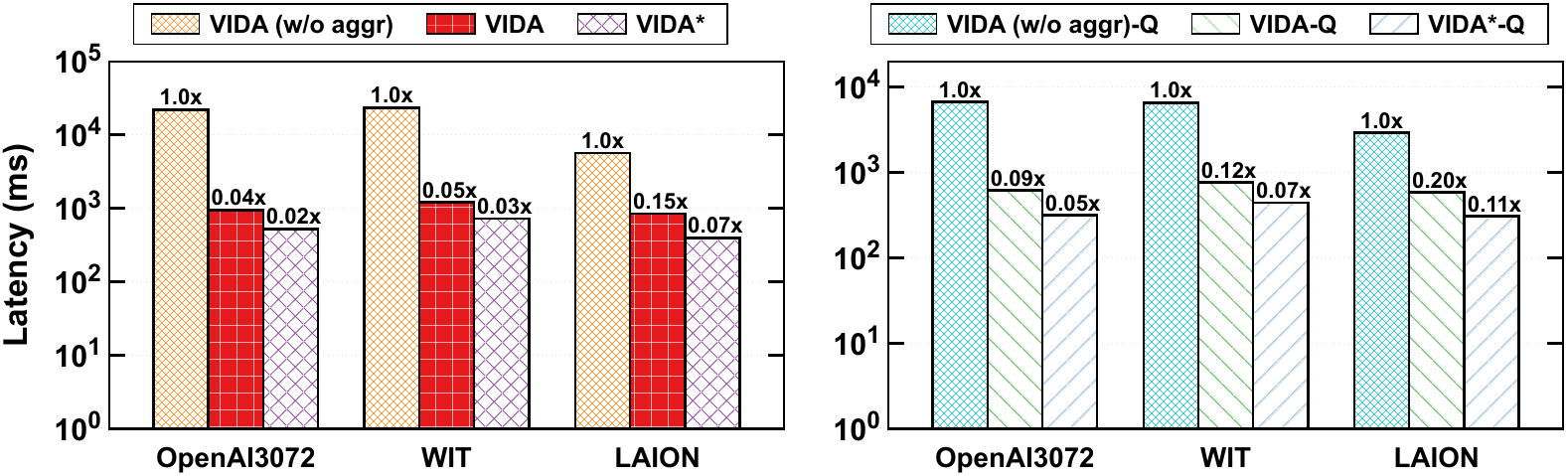}
        \caption{Ablation Study}
        \label{fig:ablation}
    \end{minipage}
\end{figure}

\subsubsection{\blue{Ablation on Core Optimizations}}
\blue{Finally, we evaluate two core optimizations: proof aggregation and the topology-aware ADS. We compare three variants. VIDA (w/o aggr) uses the same MHT structure as Strawman, but replaces raw-vector replay with individual IPA proofs for each visited node and applies batch verification~\cite{bulletproofs}. VIDA further replaces individual IPA proofs with a single aggregated proof. VIDA* further augments VIDA with \indname, our topology-aware ADS. We also include their quantization-aware versions.
As shown in \Cref{fig:ablation}, proof aggregation is essential. Since VIDA (w/o aggr) and VIDA use the same MHT structure, their gap isolates the benefit of aggregation. VIDA reduces total latency over VIDA (w/o aggr) by 91.9\% on average, while VIDA-Q reduces latency over VIDA (w/o aggr)-Q by 86.3\%. The improvement is especially significant on \textsf{OpenAI3072}, where aggregation provides 23.3$\times$ and 10.8$\times$ speedups in the non-quantized and quantization-aware settings, respectively. The gap between VIDA and VIDA* further isolates the benefit of \indname: VIDA* reduces total latency by 46.0\% over VIDA on average, and VIDA*-Q reduces latency by 46.3\% over VIDA-Q. These results confirm that aggregation removes the dominant per-node IPA bottleneck, while \indname further reduces structural authentication overhead.
}

\section{Related Work}\label{sec:related_work}

\textbf{ANN Search.} ANN search algorithms are broadly categorized into hash-based~\cite{lsh}, inverted-file-based~\cite{pq}, and graph-based methods~\cite{graphann_benchmark_sigmod}. Currently, graph-based methods are widely recognized for achieving the best trade-off between search efficiency and recall \cite{vector_database_survey}. These methods operate as a greedy best-first search on a proximity graph. Based on graph topology and edge pruning strategies, graph-based ANN methods are categorized into hierarchical (\eg, HNSW \cite{HNSW}, HVS \cite{hvs}) and flat structures (\eg, NSG \cite{nsg}, DPG \cite{dpg}). Recent research has expanded graph indexes to diverse environments, including disk-resident designs for massive datasets (\eg, DiskANN \cite{diskann}), hardware-aware layouts for GPUs (\eg, CAGRA \cite{cagra}), and metadata-filtered searches \cite{filterann}.
Despite these optimizations, their core search paradigm remains reliant on sequential graph traversals. 

\noindent
\textbf{Verifiable Query Processing}. 
Verifiable query processing over structured databases has been widely studied using authenticated data structures (ADS) \cite{integridb} or zero-knowledge proofs \cite{vsql, zksql} to ensure query integrity across various query types \cite{authenticated_keyword,vchain+,e3fs,vgq}. When extending verifiable queries to vector data, existing works mainly target exact $k$NN search. These works generally leverage spatial partitioning ADS (\eg, MR-trees~\cite{zhang2018distributed}, Voronoi diagrams~\cite{cui2023towards}) to ensure result integrity by rigorously proving that the unsearched space cannot contain better candidates. Regarding ANN search, however, this exact verification paradigm is fundamentally incompatible: high dimensionality renders spatial geometric proofs prohibitively expensive, while enforcing exactness contradicts the efficiency goal of approximate search. Consequently, verifiable ANN should shift its focus from strictly verifying the final exact result to guaranteeing \textit{execution integrity}, ensuring that the search algorithm is executed correctly and faithfully.

For graph-based ANN, ANNProof~\cite{annproof_baseline1} and Wang et al. ~\cite{graph_baseline2} use MHTs and raw-vector replay to verify HNSW and HCNNG searches, respectively. Wang et al. further reduce the number of probed nodes through HCNNG-specific optimizations. V3DB~\cite{v3db} and zkRAG~\cite{zkrag} verify IVF-PQ and HNSW retrieval, respectively, using succinct zero-knowledge proofs that hide private database and index information. However, generating such proofs incurs substantial computational overhead. For example, zkRAG reports a proof generation time of approximately 26 seconds per query on SIFT1M. In contrast, VIDA targets general graph-based ANN methods. It integrates vector commitments and aggregated IPA to authenticate distances without transmitting raw vectors, and uses topology-aware MeST to reduce structural proof overhead.
Separately, privacy-preserving ANN methods protect vector values and access patterns from service providers~\cite{compass, sanns, panther, pacmann, ppanns}.

\section{Conclusion}\label{sec:conclusion}

In this paper, we presented VIDA for efficient verifiable graph-based ANN search in outsourced vector databases. VIDA combines vector commitments with aggregated IPA proofs to authenticate distances without transmitting raw vectors. Furthermore, we developed MeST, a topology-aware Merkle spanning tree index that aligns structural verification paths with graph traversal sequences to minimize required Merkle proofs. We also adapted VIDA to quantization-aware graph search to avoid unnecessary exact-distance proofs. Our formal security analysis and experimental evaluations demonstrated that the proposed approach is secure and highly scalable. It delivers up to a \blue{6.6$\times$} speedup in end-to-end latency while reducing data transfer time by \blue{99.2\%} compared with state-of-the-art methods.

\section*{AI Usage Declaration}
Generative AI tools were used only for polishing the writing and assisting with code implementation.

\bibliographystyle{IEEEtran}
\bibliography{ref}

\begin{thebibliography}{10}
\providecommand{\url}[1]{#1}
\csname url@samestyle\endcsname
\providecommand{\newblock}{\relax}
\providecommand{\bibinfo}[2]{#2}
\providecommand{\BIBentrySTDinterwordspacing}{\spaceskip=0pt\relax}
\providecommand{\BIBentryALTinterwordstretchfactor}{4}
\providecommand{\BIBentryALTinterwordspacing}{\spaceskip=\fontdimen2\font plus
\BIBentryALTinterwordstretchfactor\fontdimen3\font minus
  \fontdimen4\font\relax}
\providecommand{\BIBforeignlanguage}[2]{{%
\expandafter\ifx\csname l@#1\endcsname\relax
\typeout{** WARNING: IEEEtran.bst: No hyphenation pattern has been}%
\typeout{** loaded for the language `#1'. Using the pattern for}%
\typeout{** the default language instead.}%
\else
\language=\csname l@#1\endcsname
\fi
#2}}
\providecommand{\BIBdecl}{\relax}
\BIBdecl

\bibitem{Milvus}
``{Milvus},'' \url{https://milvus.io}, [Accessed: 2026-6-1].

\bibitem{qdrant}
``{Qdrant},'' \url{https://qdrant.tech}, [Accessed: 2026-6-1].

\bibitem{weaviate}
``{Weaviate},'' \url{https://weaviate.io}, [Accessed: 2026-6-1].

\bibitem{rag}
P.~Lewis, E.~Perez, A.~Piktus, F.~Petroni, V.~Karpukhin, N.~Goyal,
  H.~K{\"u}ttler, M.~Lewis, W.-t. Yih, T.~Rockt{\"a}schel \emph{et~al.},
  ``Retrieval-augmented generation for knowledge-intensive {NLP} tasks,''
  \emph{Proc. Adv. Neural Inf. Process. Syst.}, vol.~33, pp. 9459--9474, 2020.

\bibitem{recommender}
S.~Zhang, L.~Yao, A.~Sun, and Y.~Tay, ``Deep learning based recommender system:
  A survey and new perspectives,'' \emph{{ACM} Comput. Surv.}, vol.~52, no.~1,
  pp. 1--38, 2019.

\bibitem{Dense_Passage_Retrieval}
V.~Karpukhin, B.~Oguz, S.~Min, P.~S. Lewis, L.~Wu, S.~Edunov, D.~Chen, and
  W.-t. Yih, ``Dense passage retrieval for open-domain question answering,'' in
  \emph{Proc. Conf. Empir. Methods Nat. Lang. Process. ({EMNLP})}, 2020, pp.
  6769--6781.

\bibitem{lsh}
Y.~Liu, J.~Cui, Z.~Huang, H.~Li, and H.~T. Shen, ``{SK-LSH}: An efficient index
  structure for approximate nearest neighbor search,'' \emph{Proc. {VLDB}
  Endow.}, vol.~7, no.~9, pp. 745--756, 2014.

\bibitem{pq}
H.~Jegou, M.~Douze, and C.~Schmid, ``Product quantization for nearest neighbor
  search,'' \emph{{IEEE} Trans. Pattern Anal. Mach. Intell.}, vol.~33, no.~1,
  pp. 117--128, 2010.

\bibitem{HNSW}
Y.~A. Malkov and D.~A. Yashunin, ``Efficient and robust approximate nearest
  neighbor search using hierarchical navigable small world graphs,''
  \emph{{IEEE} Trans. Pattern Anal. Mach. Intell.}, vol.~42, no.~4, pp.
  824--836, 2018.

\bibitem{nsg}
C.~Fu, C.~Xiang, C.~Wang, and D.~Cai, ``Fast approximate nearest neighbor
  search with the navigating spreading-out graph,'' \emph{Proc. {VLDB} Endow.},
  vol.~12, no.~5, pp. 461--474, 2019.

\bibitem{vector_database_survey}
J.~J. Pan, J.~Wang, and G.~Li, ``Survey of vector database management
  systems,'' \emph{{VLDB} J.}, vol.~33, no.~5, pp. 1591--1615, 2024.

\bibitem{song2026vector2}
Y.~Song, X.~Zhou, C.~S. Jensen, and J.~Xu, ``Vector search for the future: From
  memory-resident, static heterogeneous storage, to cloud-native
  architectures,'' in \emph{Proc. {ACM} Manage. Data}, 2026.

\bibitem{BlendHouse}
Z.~Niu, X.~Tian, X.~Peng, and X.~Chen, ``{BlendHouse}: A cloud-native vector
  database system in {ByteHouse},'' in \emph{Proc. {IEEE} Int. Conf. Data Eng.
  ({ICDE})}.\hskip 1em plus 0.5em minus 0.4em\relax IEEE, 2025, pp. 4332--4345.

\bibitem{upreti2025cost}
N.~Upreti, H.~V. Simhadri, H.~S. Sundar, K.~Sundaram, S.~Boshra,
  B.~Perumalswamy, S.~Atri, M.~Chisholm, R.~R. Singh, G.~Yang \emph{et~al.},
  ``Cost-effective, low latency vector search with azure cosmos db,''
  \emph{Proc. {VLDB} Endow.}, vol.~18, no.~12, pp. 5166--5183, 2025.

\bibitem{integridb}
Y.~Zhang, J.~Katz, and C.~Papamanthou, ``{IntegriDB}: Verifiable {SQL} for
  outsourced databases,'' in \emph{Proc. {ACM SIGSAC} Conf. Comput. Commun.
  Secur. (CCS)}, 2015, pp. 1480--1491.

\bibitem{vsql}
Y.~Zhang, D.~Genkin, J.~Katz, D.~Papadopoulos, and C.~Papamanthou, ``{vSQL}:
  Verifying arbitrary {SQL} queries over dynamic outsourced databases,'' in
  \emph{Proc. {IEEE} Symp. Secur. Privacy ({SP})}.\hskip 1em plus 0.5em minus
  0.4em\relax IEEE, 2017, pp. 863--880.

\bibitem{zksql}
X.~Li, C.~Weng, Y.~Xu, X.~Wang, and J.~Rogers, ``{ZKSQL}: Verifiable and
  efficient query evaluation with zero-knowledge proofs,'' \emph{Proc. {VLDB}
  Endow.}, vol.~16, no.~8, 2023.

\bibitem{annproof_baseline1}
L.~Lu, Z.~Wen, Y.~Yuan, Q.~He, J.~Chen, and Z.~Liu, ``{ANNProof}: Building a
  verifiable and efficient outsourced approximate nearest neighbor search
  system on blockchain,'' \emph{Future Gener. Comput. Syst.}, vol. 156, pp.
  206--220, 2024.

\bibitem{graph_baseline2}
C.~Wang, J.~Zhang, X.~Liu, K.~Wei, and B.~Feng, ``Verifiable graph-based
  approximate nearest neighbor search,'' in \emph{Proc. Int. Conf. Adv. Data
  Min. Appl.}\hskip 1em plus 0.5em minus 0.4em\relax Springer, 2024, pp. 3--17.

\bibitem{merkletree}
R.~C. Merkle, ``A certified digital signature,'' in \emph{Proc.
  {CRYPTO}}.\hskip 1em plus 0.5em minus 0.4em\relax Springer, 1989, pp.
  218--238.

\bibitem{openai3072}
``{OpenAI3072},''
  \url{https://huggingface.co/datasets/Qdrant/dbpedia-entities-openai3-text-embedding-3-large-3072-1M},
  [Accessed: 2026-6-1].

\bibitem{linearproof}
C.~Gentry, S.~Halevi, and V.~Lyubashevsky, ``Practical non-interactive publicly
  verifiable secret sharing with thousands of parties,'' in \emph{Proc.
  {EUROCRYPT}}.\hskip 1em plus 0.5em minus 0.4em\relax Springer, 2022, pp.
  458--487.

\bibitem{springproofs}
J.~Zhang, M.~Su, X.~Liu, and G.~Wang, ``Springproofs: Efficient inner product
  arguments for vectors of arbitrary length,'' in \emph{Proc. {IEEE} Symp.
  Secur. Privacy ({SP})}.\hskip 1em plus 0.5em minus 0.4em\relax IEEE, 2024,
  pp. 3147--3164.

\bibitem{pedersen}
T.~P. Pedersen, ``Non-interactive and information-theoretic secure verifiable
  secret sharing,'' in \emph{Proc. {CRYPTO}}.\hskip 1em plus 0.5em minus
  0.4em\relax Springer, 1991, pp. 129--140.

\bibitem{10.1007/978-3-642-38980-1_32}
R.~Henry and I.~Goldberg, ``Batch proofs of partial knowledge,'' in \emph{Proc.
  Appl. Cryptogr. Netw. Secur.}\hskip 1em plus 0.5em minus 0.4em\relax
  Springer, 2013, pp. 502--517.

\bibitem{fs}
M.~Bellare and P.~Rogaway, ``Random oracles are practical: A paradigm for
  designing efficient protocols,'' in \emph{Proc. {ACM SIGSAC} Conf. Comput.
  Commun. Secur. (CCS)}, 1993, pp. 62--73.

\bibitem{sq}
W.~Zhou, Y.~Lu, H.~Li, and Q.~Tian, ``Scalar quantization for large scale image
  search,'' in \emph{Proc. {ACM} Int. Conf. Multimedia ({MM})}.\hskip 1em plus
  0.5em minus 0.4em\relax ACM, 2012, pp. 169--178.

\bibitem{rabitq}
J.~Gao and C.~Long, ``{RaBitQ}: Quantizing high-dimensional vectors with a
  theoretical error bound for approximate nearest neighbor search,''
  \emph{Proc. {ACM} Manage. Data}, vol.~2, no.~3, pp. 1--27, 2024.

\bibitem{bulletproofs}
B.~B{\"u}nz, J.~Bootle, D.~Boneh, A.~Poelstra, P.~Wuille, and G.~Maxwell,
  ``Bulletproofs: Short proofs for confidential transactions and more,'' in
  \emph{Proc. {IEEE} Symp. Secur. Privacy ({SP})}.\hskip 1em plus 0.5em minus
  0.4em\relax IEEE, 2018, pp. 315--334.

\bibitem{forkinglemma}
J.~Bootle, A.~Cerulli, P.~Chaidos, J.~Groth, and C.~Petit, ``Efficient
  zero-knowledge arguments for arithmetic circuits in the discrete log
  setting,'' in \emph{Proc. Annu. Int. Conf. Theory Appl. Cryptogr.
  Tech.}\hskip 1em plus 0.5em minus 0.4em\relax Springer, 2016, pp. 327--357.

\bibitem{graph_benchmark_21}
M.~Wang, X.~Xu, Q.~Yue, and Y.~Wang, ``A comprehensive survey and experimental
  comparison of graph-based approximate nearest neighbor search,'' \emph{Proc.
  {VLDB} Endow.}, vol.~14, no.~11, pp. 1964--1978, 2021.

\bibitem{WIT}
K.~Srinivasan, K.~Raman, J.~Chen, M.~Bendersky, and M.~Najork, ``{WIT}:
  Wikipedia-based image text dataset for multimodal multilingual machine
  learning,'' in \emph{Proc. {ACM SIGIR} Int. Conf. Res. Dev. Inf. Retr.},
  2021, pp. 2443--2449.

\bibitem{LAION}
C.~Schuhmann, R.~Vencu, R.~Beaumont, R.~Kaczmarczyk, C.~Mullis, A.~Katta,
  T.~Coombes, J.~Jitsev, and A.~Komatsuzaki, ``{LAION-400M}: Open dataset of
  clip-filtered 400 million image-text pairs,'' \emph{arXiv preprint
  arXiv:2111.02114}, 2021.

\bibitem{Wikipedia}
``{Wikipedia},''
  \url{https://huggingface.co/datasets/Cohere/wikipedia-2023-11-embed-multilingual-v3},
  [Accessed: 2026-6-1].

\bibitem{Msmarco}
``{Msmarco},''
  \url{https://huggingface.co/datasets/Cohere/msmarco-v2.1-embed-english-v3},
  [Accessed: 2026-6-1].

\bibitem{bulletproofs_library}
``{Bulletproofs},'' \url{https://github.com/zkcrypto/bulletproofs}, [Accessed:
  2026-6-1].

\bibitem{toxiproxy}
``{Toxiproxy},'' \url{https://github.com/Shopify/toxiproxy}, [Accessed:
  2026-6-1].

\bibitem{yang2022mobile}
X.~Yang, H.~Lin, Z.~Li, F.~Qian, X.~Li, Z.~He, X.~Wu, X.~Wang, Y.~Liu, Z.~Liao,
  D.~Hu, and T.~Xu, ``Mobile access bandwidth in practice: Measurement,
  analysis, and implications,'' in \emph{Proc. {ACM SIGCOMM} Conf.}, 2022, pp.
  114--128.

\bibitem{graphann_benchmark_sigmod}
I.~Azizi, K.~Echihabi, and T.~Palpanas, ``Graph-based vector search: An
  experimental evaluation of the state-of-the-art,'' \emph{Proc. {ACM} Manage.
  Data}, vol.~3, no.~1, pp. 1--31, 2025.

\bibitem{hvs}
K.~Lu, M.~Kudo, C.~Xiao, and Y.~Ishikawa, ``{HVS}: Hierarchical graph structure
  based on voronoi diagrams for solving approximate nearest neighbor search,''
  \emph{Proc. {VLDB} Endow.}, vol.~15, no.~2, pp. 246--258, 2021.

\bibitem{dpg}
W.~Li, Y.~Zhang, Y.~Sun, W.~Wang, M.~Li, W.~Zhang, and X.~Lin, ``Approximate
  nearest neighbor search on high dimensional data—experiments, analyses, and
  improvement,'' \emph{{IEEE} Trans. Knowl. Data Eng.}, vol.~32, no.~8, pp.
  1475--1488, 2019.

\bibitem{diskann}
S.~Jayaram~Subramanya, F.~Devvrit, H.~V. Simhadri, R.~Krishnawamy, and
  R.~Kadekodi, ``Diskann: Fast accurate billion-point nearest neighbor search
  on a single node,'' \emph{Proc. Adv. Neural Inf. Process. Syst.}, vol.~32,
  2019.

\bibitem{cagra}
H.~Ootomo, A.~Naruse, C.~Nolet, R.~Wang, T.~Feher, and Y.~Wang, ``Cagra: Highly
  parallel graph construction and approximate nearest neighbor search for
  {GPUs},'' in \emph{Proc. {IEEE} Int. Conf. Data Eng. ({ICDE})}.\hskip 1em
  plus 0.5em minus 0.4em\relax IEEE, 2024, pp. 4236--4247.

\bibitem{filterann}
Y.~Lin, K.~Zhang, Z.~He, Y.~Jing, and X.~S. Wang, ``Survey of filtered
  approximate nearest neighbor search over the vector-scalar hybrid data,''
  \emph{arXiv preprint arXiv:2505.06501}, 2025.

\bibitem{authenticated_keyword}
C.~Zhang, C.~Xu, H.~Wang, J.~Xu, and B.~Choi, ``Authenticated keyword search in
  scalable hybrid-storage blockchains,'' in \emph{Proc. {IEEE} Int. Conf. Data
  Eng. ({ICDE})}.\hskip 1em plus 0.5em minus 0.4em\relax IEEE, 2021, pp.
  996--1007.

\bibitem{vchain+}
H.~Wang, C.~Xu, C.~Zhang, J.~Xu, Z.~Peng, and J.~Pei, ``{vChain+}: Optimizing
  verifiable blockchain boolean range queries,'' in \emph{Proc. {IEEE} Int.
  Conf. Data Eng. ({ICDE})}.\hskip 1em plus 0.5em minus 0.4em\relax IEEE, 2022,
  pp. 1927--1940.

\bibitem{e3fs}
P.~Sun, L.~Zhang, J.~Liu, C.~Tang, and J.~Wang, ``{E$^3$FS}: Efficient, secure,
  and verifiable fuzzy search with data updates in hybrid-storage
  blockchains,'' in \emph{Proc. {IEEE} Int. Conf. Data Eng. ({ICDE})}.\hskip
  1em plus 0.5em minus 0.4em\relax IEEE, 2025, pp. 3890--3903.

\bibitem{vgq}
Z.~Yao, T.~Li, J.~Xin, Y.~Li, C.~Wang, Z.~Wang, D.~Srivastava, and C.~S.
  Jensen, ``{VGQ}: Enabling verifiable graph queries on blockchain systems,''
  in \emph{Proc. {IEEE} Int. Conf. Data Eng. ({ICDE})}.\hskip 1em plus 0.5em
  minus 0.4em\relax IEEE, 2025, pp. 3602--3614.

\bibitem{zhang2018distributed}
C.~Zhang, C.~Xu, J.~Xu, and B.~Choi, ``Distributed {kNN} query
  authentication,'' in \emph{Proc. {IEEE} Int. Conf. Mobile Data Manage.
  ({MDM})}.\hskip 1em plus 0.5em minus 0.4em\relax IEEE, 2018, pp. 167--176.

\bibitem{cui2023towards}
N.~Cui, K.~Qian, T.~Cai, J.~Li, X.~Yang, J.~Cui, and H.~Zhong, ``Towards
  multi-user, secure, and verifiable {kNN} query in cloud database,''
  \emph{{IEEE} Trans. Knowl. Data Eng.}, vol.~35, no.~9, pp. 9333--9349, 2023.

\bibitem{v3db}
Z.~Qiu, W.~Qu, J.~Zhang, and B.~Yuan, ``{V3DB}: Audit-on-demand zero-knowledge
  proofs for verifiable vector search over committed snapshots,'' \emph{Proc.
  {VLDB} Endow.}, vol.~19, no.~10, pp. 2604--2616, 2026.

\bibitem{zkrag}
\BIBentryALTinterwordspacing
Y.~Jiang, X.~Yang, X.~Liu, Y.~Guo, and J.~Zhang, ``{zkRAG}: Efficiently proving
  {RAG} retrieval in zero knowledge,'' \emph{Cryptology ePrint Archive}, no.
  2026/709, 2026. [Online]. Available: \url{https://eprint.iacr.org/2026/709}
\BIBentrySTDinterwordspacing

\bibitem{compass}
J.~Zhu, L.~Patel, M.~Zaharia, and R.~A. Popa, ``Compass: Encrypted semantic
  search with high accuracy,'' in \emph{Proc. {USENIX} Symp. Oper. Syst. Des.
  Implement. ({OSDI})}, 2025, pp. 915--938.

\bibitem{sanns}
H.~Chen, I.~Chillotti, Y.~Dong, O.~Poburinnaya, I.~Razenshteyn, and M.~S.
  Riazi, ``{SANNS}: Scaling up secure approximate k-nearest neighbors search,''
  in \emph{Proc. {USENIX} Secur. Symp.}, 2020, pp. 2111--2128.

\bibitem{panther}
J.~Li, Z.~Huang, M.~Zhang, C.~Hong, J.~Liu, T.~Wei, and W.~Chen, ``Panther:
  Private approximate nearest neighbor search in the single server setting,''
  in \emph{Proc. {ACM SIGSAC} Conf. Comput. Commun. Secur. (CCS)}, 2025, pp.
  365--379.

\bibitem{pacmann}
M.~Zhou, E.~Shi, and G.~Fanti, ``Pacmann: Efficient private approximate nearest
  neighbor search,'' \emph{Cryptology ePrint Archive}, 2024.

\bibitem{ppanns}
Y.~Liu, Y.~Zhang, J.~Xie, H.~Li, J.~X. Yu, and J.~Cui, ``Privacy-preserving
  approximate nearest neighbor search on high-dimensional data,'' in
  \emph{Proc. {IEEE} Int. Conf. Data Eng. ({ICDE})}.\hskip 1em plus 0.5em minus
  0.4em\relax IEEE, 2025, pp. 3017--3029.

\end{thebibliography}

\iftechreport

\appendices

\section{\indname for Multi-layered Graphs}\label{append-mest-multilayer}

In graph-based ANN search, a prominent class of algorithms employs multi-layered graph structures~\cite{HNSW}, which effectively function as skip lists to accelerate query processing. These indexes organize the graph into a hierarchy, enabling the query to navigate from sparser upper layers all the way down to the dense base layer. To further exploit this hierarchical topology, we extend MeST with a layer-aware design that aligns the Merkle spanning tree with the existing graph layers.

\textbf{ADS Construction.}
To construct a unified spanning tree across multiple layers, the DO first consolidates the subgraphs from all layers into a single unified graph. By flattening the hierarchical structure, the DO can invoke \cref{alg:obfs} on the resulting graph to build a distance-aware spanning tree. This flattening step ensures that the spanning tree captures the skip-list jumps provided by the original hierarchical index. Since the ANN search may traverse neighbors within a specific layer at any given time, we introduce a Multi-layered Graph Node ADS to maintain this structural granularity.
\begin{definition}[Multi-layered Graph Node ADS]
For each graph node $v_i$ in the multi-layered index $\mathcal{I}$, its corresponding ADS $g_i$ consists of
(i) $ID_i$: the identifier of the current node $v_i$;
(ii) multiple layer-specific neighbors: for each layer $0 \le l \le L_i$, a set $g_i^{(l)}=\{(ID_j, C_j, \|\mathbf{x}_j\|)\}_{\forall v_j \in Adj_i^{(l)}}$ containing the identifiers, commitments, and norms of the immediate neighbors of $v_i$ at layer $l$. Here, $Adj_i^{(l)}$ denotes the immediate neighbors at that layer. 
The hash of $g_i$ is defined as
$
H_i^{g} = \mathbb{H}(ID_i \concat H_i^{(0)} \concat \dotsm \concat H_i^{(L_i)})
$,
where $H_i^{(l)}$ is the hash for the neighbors at the $l$-th layer, defined as 
$
H_i^{(l)} = \mathbb{H}(g_i^{(l)}) =  \mathbb{H}(\{ ID_j \concat C_j \concat \|\mathbf{x}_j\|\}_{\forall v_j \in Adj_i^{(l)}})
$.
\end{definition}
The remainder of the ADS structure and its construction algorithm are identical to those described in \cref{sec:Mest_ADS}.

\textit{\textbf{Example}. Consider the two-layered proximity graph in \Cref{fig-mest_hi}. \indname first forms a unified graph by taking the union of edges across all layers, then builds the spanning tree over this unified graph. For example, starting from the entry point $v_1$, the spanning tree algorithm reaches node $v_4$ through the adjacency list of $v_1$ on layer $L_1$, and reaches nodes $v_5$, $v_6$, and $v_{12}$ through the adjacency list of $v_1$ on layer $L_0$. Based on the spanning tree, \indname is constructed in a layer-aware manner, where the graph ADS of each node is organized into $g_i^{(1)}$ and $g_i^{(0)}$. For instance, $g_5^{(1)}$ contains $(ID_1,C_1,\|\mathbf{x}_1\|)$ and $(ID_4,C_4,\|\mathbf{x}_4\|)$, corresponding to its neighbors $v_1$ and $v_4$ on layer $L_1$, while $v_{12}$ only has $g_{12}^{(0)}$ because it does not appear on layer $L_1$.}

\begin{figure}[t]
    \centering
    \includegraphics[width=0.85\linewidth]{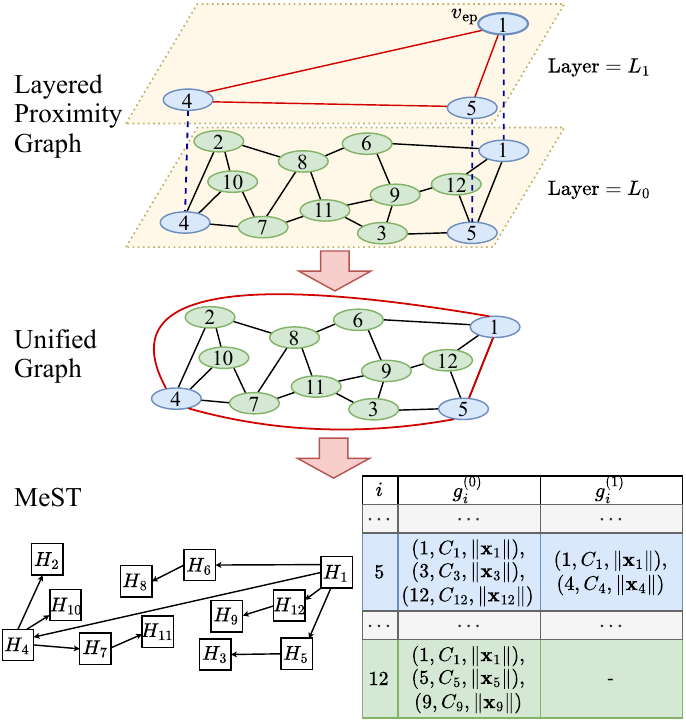}
    \caption{An Example on a Multi-layered Graph}
    \label{fig-mest_hi}
\end{figure}

\textbf{Verifiable Search.}
In the multi-layered graph index, the search state is determined not only by the visited node, but also by the layer at which the node is accessed. To avoid disclosing redundant neighborhood information, we adopt a layer-specific disclosure strategy. Concretely, if a node $v_i$ is accessed at layer $l$ during the search, the SP reveals only the neighborhood payload $g_i^{(l)}$ for that layer.
For any unaccessed layer $l'$, the VO contains only the corresponding digest $H_i^{(l')}$ rather than the full neighbor information. Therefore, the VO materializes only the layer-specific data actually used during the search, while the remaining layers are authenticated via their hashes.
During verification, the client replays the designated ANN algorithm from the top layer all the way down to the base layer. If the replay attempts to access the neighborhood information of a node $v_i$ on a layer whose payload is not disclosed in the VO, the verification fails immediately. In this way, \indname avoids transmitting unused layer-specific graph information.

\section{Soundness of IPA Aggregation}\label{append-soundness}
\label{sec:security_aggregation}

Our aggregation mechanism relies on the computational knowledge soundness of the underlying IPA.
Intuitively, this property guarantees that a prover cannot produce a valid proof without actually knowing the corresponding witness (\ie, the vector value).

\begin{lemma}[Soundness of Single IPA]
\label{lemma:single-ipa}
The single IPA protocol used in our scheme satisfies \textit{computational knowledge soundness}. That is, for any PPT adversary $\textsf{Adv}$ producing a valid proof with non-negligible probability, there exists a PPT extractor $\mathcal{E}_{IPA}$ that extracts the witness $\mathbf{x}_i$ in the relation $\langle \mathbf{x}_i, \mathbf{x}_q \rangle = IP_i$.
\end{lemma}

\begin{proof}
This claim follows from the soundness of the underlying IPA protocol, such as Bulletproofs~\cite{bulletproofs}.
\end{proof}

\AggSoundness*

\begin{proof}
The proof proceeds by constructing an extractor $\mathcal{E}$ that utilizes the adversary $\textsf{Adv}$ as a black box.

\textbf{Transcript Generation via Rewinding.}
The aggregation protocol involves a random challenge scalar $r$ derived via Fiat-Shamir. Let $n$ be the number of aggregated instances. The extractor $\mathcal{E}$ runs $\textsf{Adv}$ to obtain a valid transcript with an initial challenge $r_1$. By applying the General Forking Lemma \cite{forkinglemma}, $\mathcal{E}$ rewinds $\textsf{Adv}$ to the point of challenge generation and re-runs it multiple times with distinct random challenges $r_2, \dots, r_n$.
Assuming $\textsf{Adv}$ succeeds with non-negligible probability $\epsilon$, $\mathcal{E}$ obtains $n$ valid transcripts corresponding to a set of distinct challenges $\{r_1, r_2, \dots, r_n\}$. Note that while extending to $n$ branches introduces a polynomial reduction loss, the success probability remains non-negligible for polynomial $n$.

\textbf{Extraction of Aggregated Witness.}
For each transcript $\ell \in \{1, \dots, n\}$, the underlying proof is a valid single IPA for the aggregated statement defined by the challenge $r_\ell$:
\begin{equation}
    C_{agg}^{(\ell)} = \sum_{i=1}^n (r_\ell)^i \cdot C_i, \quad IP_{agg}^{(\ell)} = \sum_{i=1}^n (r_\ell)^i \cdot IP_i
\end{equation}
Since the underlying IPA protocol satisfies computational knowledge soundness, for each transcript $\ell$, we can invoke its extractor $\mathcal{E}_{IPA}$ to extract a witness vector $\hat{\mathbf{x}}^{(\ell)}$ satisfying:
\begin{equation}
    \funcname{Com}(\hat{\mathbf{x}}^{(\ell)}) = C_{agg}^{(\ell)} \quad \text{and} \quad \langle \hat{\mathbf{x}}^{(\ell)}, \mathbf{x}_q \rangle = IP_{agg}^{(\ell)}
\end{equation}
Let $\{\mathbf{x}_i\}_{i=1}^n$ be the actual witnesses. We now show how to compute them. The relationship between the extracted aggregated witnesses and the individual witnesses can be modeled as a system of linear equations:
\begin{equation}
    \begin{pmatrix}
    \hat{\mathbf{x}}^{(1)} \\ \hat{\mathbf{x}}^{(2)} \\ \vdots \\ \hat{\mathbf{x}}^{(n)}
    \end{pmatrix}
    =
    \begin{pmatrix}
    r_1 & r_1^2 & \dots & r_1^n \\
    r_2 & r_2^2 & \dots & r_2^n \\
    \vdots & \vdots & \ddots & \vdots \\
    r_n & r_n^2 & \dots & r_n^n
    \end{pmatrix}
    \begin{pmatrix}
    \mathbf{x}'_1 \\ \mathbf{x}'_2 \\ \vdots \\ \mathbf{x}'_n
    \end{pmatrix}
\end{equation}
The matrix $\mathbf{M}$ is a Vandermonde matrix, which is invertible provided all $r_\ell$ are distinct and non-zero. The extractor $\mathcal{E}$ computes the unique solution $\{\mathbf{x}'_i\}_{i=1}^n = \mathbf{M}^{-1} \cdot \{\hat{\mathbf{x}}^{(\ell)}\}_{\ell=1}^n$.

Due to the homomorphic property of the commitment scheme, the value $\sum_{i=1}^n (r_\ell)^i \mathbf{x}'_i$ is a valid opening for $C_{agg}^{(\ell)}$. Under the computational binding property of the Pedersen commitment guaranteed by the Discrete Logarithm assumption, it is computationally infeasible for the adversary to open $C_{agg}^{(\ell)}$ to a value different from the linear combination of the original openings. Therefore, except with negligible probability, the solved values must be the true witnesses committed in $C_i$. Thus, we have $\mathbf{x}_i = \mathbf{x}'_i$.

\textbf{Polynomial Analysis.}
Having extracted the valid witnesses $\mathbf{x}_i$ for the commitments, we now verify they satisfy the inner product relations. For every $\ell \in \{1, \dots, n\}$, witness extraction gives:
\begin{equation}
    \langle \hat{\mathbf{x}}^{(\ell)}, \mathbf{x}_q \rangle = IP_{agg}^{(\ell)}
\end{equation}
Substituting the extracted linear combinations:
\begin{equation}
    \left\langle \sum_{i=1}^n (r_\ell)^i \mathbf{x}_i, \mathbf{x}_q \right\rangle = \sum_{i=1}^n (r_\ell)^i IP_i
\end{equation}
Since $\mathbf{x}_q$ is fixed and public, the inner product operator $\langle \cdot, \mathbf{x}_q \rangle$ is linear. We can rewrite the equation as:
\begin{equation}
    \begin{gathered}
    \sum_{i=1}^n (r_\ell)^i \langle \mathbf{x}_i, \mathbf{x}_q \rangle = \sum_{i=1}^n (r_\ell)^i IP_i \\
    \implies \sum_{i=1}^n (r_\ell)^i \left( \langle \mathbf{x}_i, \mathbf{x}_q \rangle - IP_i \right) = 0
    \end{gathered}
\end{equation}
Let $\Delta_i = \langle \mathbf{x}_i, \mathbf{x}_q \rangle - IP_i$. This defines a homogeneous linear system $\mathbf{M} \cdot \mathbf{\Delta} = \mathbf{0}$, where $\mathbf{\Delta} = (\Delta_1, \dots, \Delta_n)^T$ and $\mathbf{M}$ is the same invertible Vandermonde matrix used for extraction.
\begin{equation}
    \begin{pmatrix}
    r_1 & r_1^2 & \dots & r_1^n \\
    r_2 & r_2^2 & \dots & r_2^n \\
    \vdots & \vdots & \ddots & \vdots \\
    r_n & r_n^2 & \dots & r_n^n
    \end{pmatrix}
    \begin{pmatrix}
    \Delta_1 \\ \Delta_2 \\ \vdots \\ \Delta_n
    \end{pmatrix}
    =
    \begin{pmatrix}
    0 \\ 0 \\ \vdots \\ 0
    \end{pmatrix}
\end{equation}
Since $\mathbf{M}$ is non-singular, the only solution is the trivial solution:
\begin{equation}
    \Delta_1 = \Delta_2 = \dots = \Delta_n = 0
\end{equation}
This implies that $\langle \mathbf{x}_i, \mathbf{x}_q \rangle - IP_i = 0$, and thus $\langle \mathbf{x}_i, \mathbf{x}_q \rangle = IP_i$ for all $i \in \{1, \dots, n\}$.

We have constructed an extractor $\mathcal{E}$ that succeeds in extracting valid witnesses for all constituent instances whenever the adversary succeeds. This proves that the aggregation protocol satisfies computational knowledge soundness.
\end{proof}

\section{Security of \schemename}\label{append-security}

\VIDASecurity*

\begin{proof}
Let \textsf{Adv} be a PPT adversary that wins the experiment in Definition~\ref{def:security} with non-negligible probability. 
Namely, given a database $\mathcal{D}$ and a query $Q$, \textsf{Adv} outputs a query result $R_{\textsf{Real}}$ and a verification object VO such that
$\funcname{Verify}(Q, R_{\textsf{Real}}, \text{VO}) = \texttt{accept}$,
yet
$R_{\textsf{Real}} \neq R_{\textsf{Ideal}}$,
where $R_{\textsf{Ideal}}$ is the output of the designated deterministic ANN algorithm faithfully executed on the authentic index generated from $\mathcal{D}$.
\blue{For the quantization-aware instantiation, the designated algorithm is the deterministic quantization-aware graph search designated by the DO.}

We show that such an adversary must break at least one of the underlying cryptographic assumptions with non-negligible probability.

\textbf{Integrity of the Local Subindex.}
During verification, the client reconstructs a local subgraph $\mathcal{I}'$ from the data carried in the VO. 
Let $\mathcal{I}$ denote the authentic ANN index generated during the setup phase. 
We analyze the integrity of $\mathcal{I}'$ from the following three aspects.

\begin{itemize}
    \item \textit{Structural Integrity.}
    If the structural information of any node in $\mathcal{I}'$ (\eg, node identifier or adjacency list) differs from that in $\mathcal{I}$, yet still passes Merkle-root verification against the signed root digest, then \textsf{Adv} must either forge a signature or find a collision in $\mathbb{H}$ in order to produce a valid proof for altered data.

    \item \textit{Data Integrity.}
    If the commitment $C_i$ or the authenticated norm $\|\mathbf{x}_i\|$ associated with a node $v_i$ in $\mathcal{I}'$ differs from the corresponding value in $\mathcal{I}$, while the verification still accepts, then again \textsf{Adv} must either forge a signature or find a hash collision. 
    Moreover, if \textsf{Adv} claims that a commitment $C_i$ opens to some vector $\mathbf{x}'_i \neq \mathbf{x}_i$, then this violates the computational binding property of the Pedersen commitment.

    \item \textit{Distance Integrity.}
    If the recovered distance between a node vector and the query vector differs from the true distance induced by the vector committed in $C_i$, then \textsf{Adv} must have produced a valid aggregated IPA proof for an invalid inner-product relation. 
    This contradicts the computational knowledge soundness of the underlying IPA protocol.
\end{itemize}

Therefore, except with negligible probability, the verifier reconstructs a faithful local subgraph $\mathcal{I}'$ of $\mathcal{I}$: for every node included in $\mathcal{I}'$, its authenticated structural information and its recovered distance to the query are identical to those in the authentic index.

\textbf{Integrity of the Replay Process.}
Let $f$ denote the designated deterministic ANN algorithm. 
In the Ideal Game, the trusted party outputs
$R_{\textsf{Ideal}} = f(Q,\mathcal{I})$.
In the Real Game, once the structural proof and the aggregated IPA proof are accepted, the verifier replays the same deterministic algorithm on the reconstructed subgraph $\mathcal{I}'$ and obtains a local replay result
$R' = f(Q,\mathcal{I}')$.
The verification procedure accepts only if the replay does not require any missing node and $R' = R_{\textsf{Real}}$.

We now analyze the ways in which \textsf{Adv} may try to make the Real Game output a result different from the Ideal Game.

\paragraph{Case 1: \textsf{Adv} provides an incomplete subgraph.}
Suppose \textsf{Adv} omits some node that is required by the honest execution of $f$ on the authentic index. 
Because the ANN search proceeds by graph traversal, every such node is reached through the authenticated adjacency list of some previously revealed node. 
Hence, during replay on $\mathcal{I}'$, the verifier will eventually attempt to access a missing node. 
According to the verification procedure, this causes rejection, and thus
$\textbf{Game}_{\textsf{Real}} = \bot$. Therefore, an incomplete subgraph cannot lead to a successful forgery.

\paragraph{Case 2: \textsf{Adv} provides a complete and faithful subgraph but deviates from the search protocol.}
Assume now that $\mathcal{I}'$ contains all nodes required by the honest search and is faithful to $\mathcal{I}$ on those nodes. 
Since the verifier replays the same deterministic ANN algorithm $f$ on the same authenticated structure and distances, the replay necessarily follows the honest behavior on the revealed subgraph. 
Consequently, the replay result coincides with the honest result:
$R' = R_{\textsf{Ideal}}$.

If \textsf{Adv} outputs any forged result $R_{\textsf{Real}} \neq R_{\textsf{Ideal}}$, then necessarily $R_{\textsf{Real}} \neq R' $, so the final equality check fails and the verifier rejects. 
Hence, in this case as well, \textsf{Adv} cannot make the Real Game output a value different from the Ideal Game while still being accepted.

\paragraph{Security of Topology-aware Instantiation.}
For the topology-aware instantiation, MeST replaces the standard MHT while preserving the binding between every authenticated graph-node opening and the signed root. Any modification to the revealed graph structure therefore implies either a hash collision or a signature forgery. Once the authenticated local subgraph is reconstructed, the replay argument is identical to that of the basic VIDA scheme.

\paragraph{Security of Quantization-aware Instantiation.}
\blue{For the quantization-aware instantiation, the replay process further contains two types of branches. For nodes represented by the exact-distance branch (\ie, nodes in $V_{\mathrm{exa}}$), the verifier follows the same exact-distance authentication procedure as in \schemename: the commitments, norms, and claimed inner products are authenticated by the ADS and the aggregated IPA proof, and hence the recovered exact distances are faithful to the committed vectors. Such authenticated exact distances are sufficient for replay even if a node could have been pruned by quantization. Specifically, if $\delta_j\leq\delta_{\mathrm{th}}$, then, under the stated guarantee of the quantization method, the pruning condition cannot hold for $v_j$, so the node cannot be validly pruned. Otherwise, if $\delta_j>\delta_{\mathrm{th}}$, the designated ANN update discards the node without changing the effective search state. Therefore, an unnecessary exact evaluation may change the verification branch, but it leaves the replay state and result unchanged.}

\blue{For nodes pruned by quantization (\ie, nodes in $V_{\mathrm{qua}}$), the verifier checks the pruning decision using the authenticated quantization payload. Specifically, each node opening $g_i^{op}$ is
authenticated by the Merkle proof and binds, for every exposed neighbor $v_j$, both the exact-distance payload digest $H_j^{exa}$ and the quantization payload digest $H_j^{qua}$. Therefore, the adversary cannot replace $\eta_j$ without causing a hash collision or forging the signed root. The verifier recomputes $LB_j=\funcname{QuantLB}(\eta_j,\mathbf{x}_q)$ and accepts the pruning branch only if $LB_j>\delta_{\mathrm{th}}$ at the same replay point. Under the same quantization guarantee, this pruning decision is valid, and the pruned node cannot improve the current search state. Consequently, every accepted replay step either follows the designated branch or replaces a valid pruning step with a state-preserving exact evaluation. In either case, the verifier reaches the same effective search state after processing the node and therefore obtains the same final result as the designated quantization-aware search.
}

Taken together, the aforementioned cases show that the only way for \textsf{Adv} to achieve
$(\textbf{Game}_{\textsf{Real}} \neq \bot)\ \land\ 
(\textbf{Game}_{\textsf{Real}} \neq \textbf{Game}_{\textsf{Ideal}})$
with non-negligible probability is to break at least one of the following assumptions: the security of the signature, collision resistance of $\mathbb{H}$, computational binding of the commitment scheme, or computational knowledge soundness of the IPA protocol. 
All of these events occur only with negligible probability. 
Therefore, the success probability of \textsf{Adv} is negligible, which proves the theorem.
\end{proof}

\section{Performance Results on Additional Datasets}\label{append:additional-evaluation}

We report additional results on \textsf{Wikipedia}, \textsf{Msmarco}, and \textsf{Cluster}, complementing \Cref{fig:latency_k,fig:latency_bandwidth}. All experimental settings and methods follow \Cref{sec:evaluation}. In \Cref{fig:latency_k_extra,fig:latency_bandwidth_extra}, latency is decomposed into server-side verifiable search (Server), network transfer (Transfer), and client-side verification (Client).

\begin{figure}[H]
    \centering
    \begin{minipage}{\linewidth}
        \centering
        \includegraphics[width=0.96\linewidth]{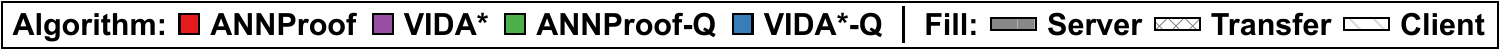}
    \end{minipage}\par\nointerlineskip\vspace{1pt}
    \includegraphics[width=0.99\linewidth,trim=0 16.5bp 0 29bp,clip]{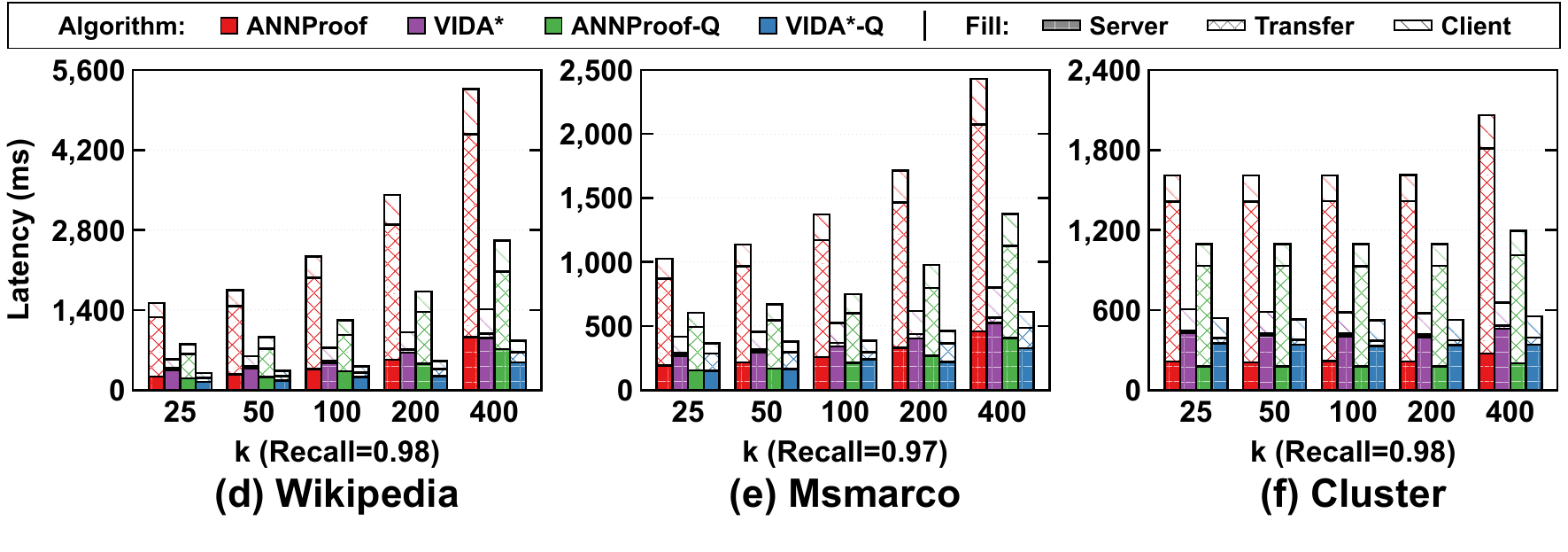}
    \caption{Latency of Varying $k$: Additional Datasets}
    \label{fig:latency_k_extra}
\end{figure}

\textbf{Impact of $k$.} We vary $k$ from 25 to 400 while maintaining target recalls for each dataset. As shown in \Cref{fig:latency_k_extra}, VIDA* and VIDA*-Q achieve lower latency than their corresponding baselines across all tested $k$ values. On \textsf{Cluster}, latency is nearly flat for $k=25$ to $200$ and increases at $k=400$, consistent with the search effort needed to maintain a fixed target recall even at small $k$.

\begin{figure}[H]
    \centering
    \begin{minipage}{\linewidth}
        \centering
        \includegraphics[width=0.96\linewidth]{pic/legend_appendix.pdf}
    \end{minipage}\par\nointerlineskip\vspace{1pt}
    \includegraphics[width=0.99\linewidth,trim=0 0 0 29bp,clip]{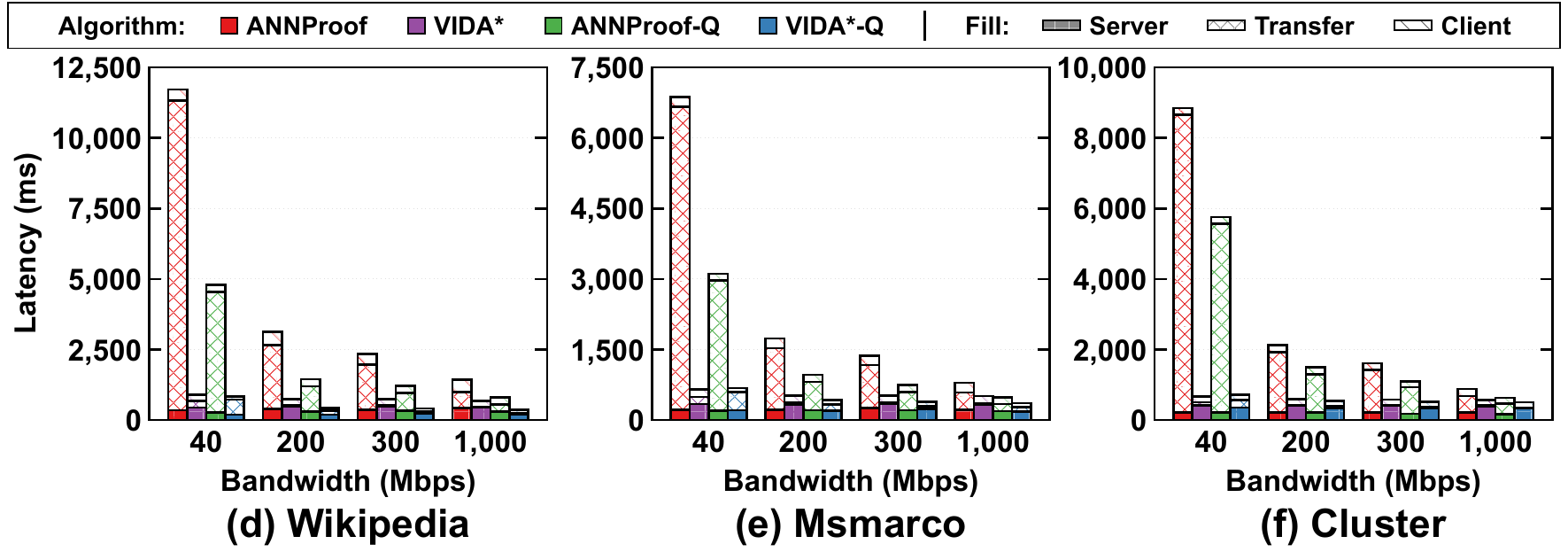}
    \caption{Latency on Different Network Settings: Additional Datasets}
    \label{fig:latency_bandwidth_extra}
\end{figure}

\textbf{Impact of Network Bandwidth.} We fix $k=100$ and use the same target recalls under four bandwidth profiles: 4G (40 Mbps), 5G (200 Mbps), Broadband (300 Mbps), and Server (1,000 Mbps). \Cref{fig:latency_bandwidth_extra} shows that VIDA* and VIDA*-Q outperform their corresponding baselines across all four bandwidth profiles, with the largest gains at 40 Mbps. Compact authentication payloads reduce the transfer cost that dominates baseline latency at this bandwidth.

\section{Additional Experimental Results}

\subsection{Offline Construction Results}\label{append:offline-construction}
We evaluate the offline overhead of index initialization across the six datasets, measuring construction time and ADS size under the settings in \Cref{sec:evaluation}.

\begin{figure}[ht]
    \centering
    \begin{minipage}{\linewidth}
        \centering
        \includegraphics[width=0.57\linewidth]{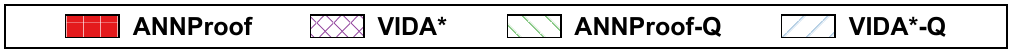}
    \end{minipage}\par\smallskip
    \begin{minipage}[b]{0.495\linewidth}
        \centering
        \includegraphics[width=0.99\linewidth]{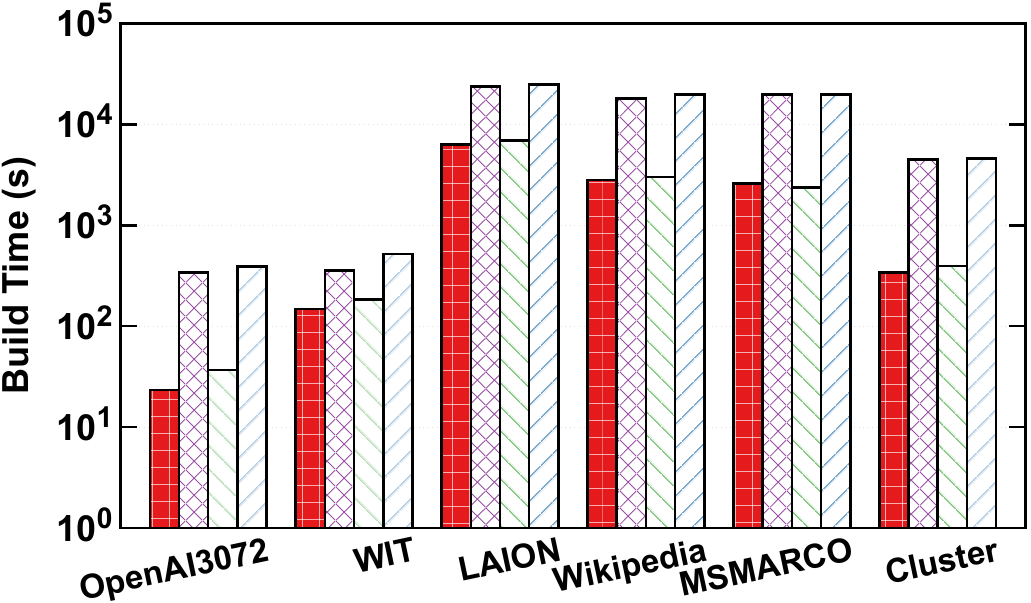}
        \caption{ADS Build Time}
        \label{fig:index_time}
    \end{minipage}\hfill%
    \begin{minipage}[b]{0.495\linewidth}
        \centering
        \includegraphics[width=0.99\linewidth]{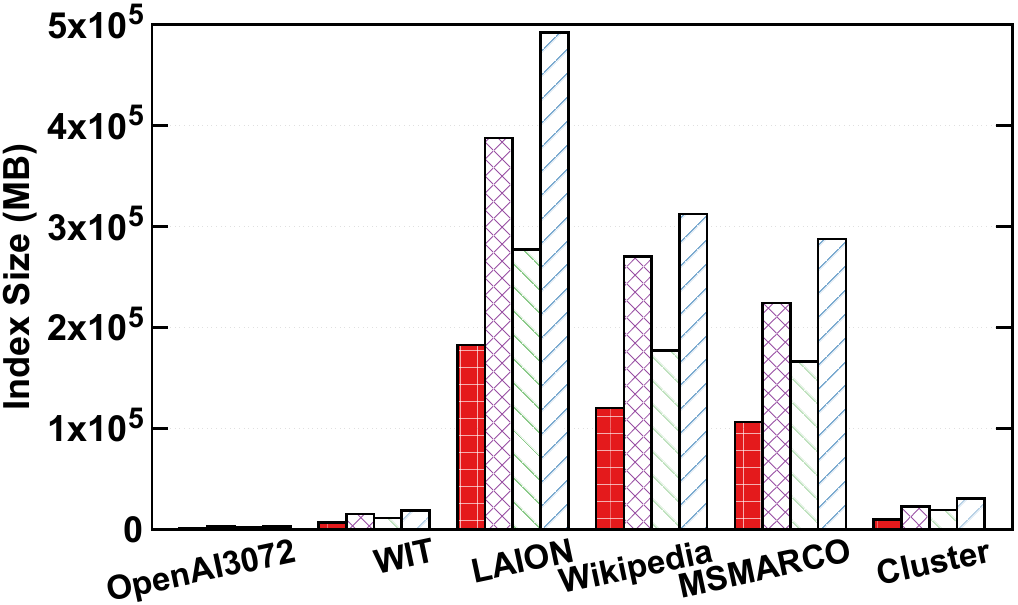}
        \caption{Index Size}
        \label{fig:index_size}
    \end{minipage}
\end{figure}

\blue{\textbf{Construction Time.} As shown in \Cref{fig:index_time}, ANNProof generally incurs the lowest construction overhead because it mainly relies on hashing. VIDA* additionally generates vector commitments and constructs the topology-aware ADS. Compared with their corresponding baselines, VIDA* and VIDA*-Q require 5.5$\times$ and 5.4$\times$ the construction time on average, respectively.}

\blue{\textbf{Storage Consumption.} \Cref{fig:index_size} shows the storage cost of the additional authentication information. VIDA* stores vector commitments and the topology-aware ADS, while VIDA*-Q further stores quantization metadata for authenticated pruning. Their ADS sizes increase by 116.1\% and 75.4\%, respectively, over their corresponding baselines.}

\blue{This offline cost is paid once by the data owner and is an acceptable trade-off for the substantial online latency and communication reductions reported in \Cref{sec:evaluation}.}

\subsection{Performance under Low Dimensions and Low Recalls}\label{append:applicability}

We further evaluate end-to-end query latency on low-dimensional \textsf{Cluster} and at lower recall targets across different values of $k$.

\begin{figure}[H]
    \centering
    \begin{minipage}{\linewidth}
        \centering
        \includegraphics[width=0.95\linewidth]{pic/legend_appendix.pdf}
    \end{minipage}\par\nointerlineskip%
    \includegraphics[width=0.5\linewidth]{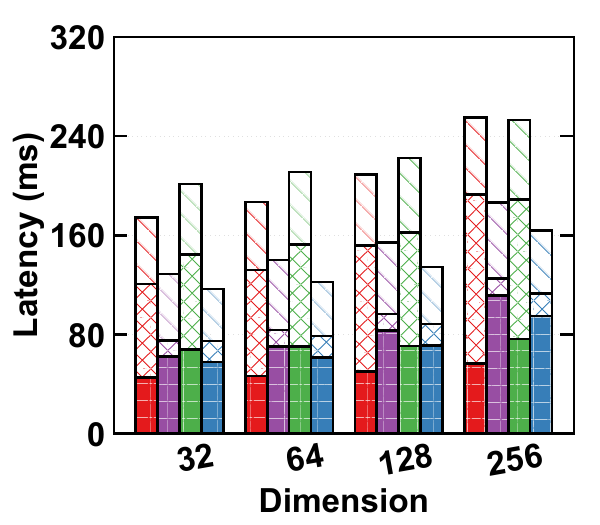}
    \caption{Latency on Low-dimensional \textsf{Cluster}}
    \label{fig:cluster_lowdim}
\end{figure}

\textbf{Performance on Low-dimensional Cluster.} Following the setup of the scalability study in \Cref{sec:scal}, we extend the dimensionality experiment on \textsf{Cluster} to $32,64,128,256$, keeping the search parameters unchanged.
\Cref{fig:cluster_lowdim} shows that VIDA* and VIDA*-Q achieve 1.34--1.37$\times$ and 1.54--1.72$\times$ speedups over their corresponding baselines, respectively, although the gains are smaller than at high dimensions. Smaller raw-vector payloads reduce the communication savings from IPA-based authentication relative to its proving and verification costs. The remaining advantage is consistent with \indname's reduction of structural authentication overhead through selective neighborhood openings and shared Merkle paths.
Compared with their non-quantized counterparts, ANNProof-Q is 15.5\% slower and VIDA*-Q is only 9.3\% faster at $d=32$. Compressing already small vectors yields limited absolute savings, while authenticating quantization metadata and pruning decisions adds overhead. Thus, quantization helps only when its computational and communication savings outweigh the added authentication cost.

\begin{figure}[H]
    \centering
    \includegraphics[width=0.9\linewidth]{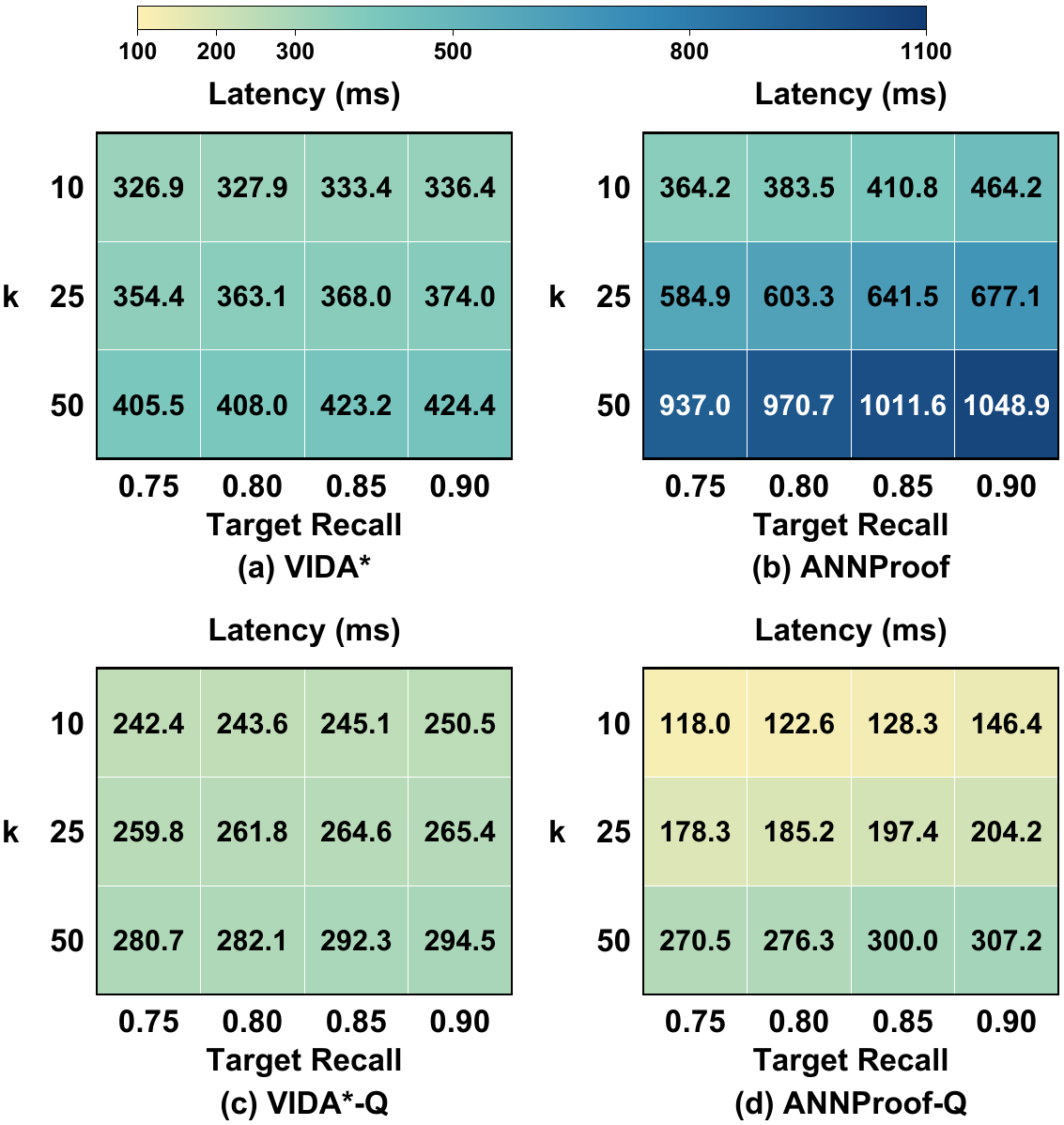}
    \caption{Latency at Lower Recall Targets on \textsf{OpenAI3072}}
    \label{fig:low_recall_matrix}
\end{figure}

\textbf{Performance under Low Selectivity and Low Recall.}
We evaluate \textsf{OpenAI3072} at $k=10,25,50$ and target recalls of $0.75,0.80,0.85,0.90$, keeping the bandwidth at 300 Mbps. \Cref{fig:low_recall_matrix} shows that VIDA* remains faster than ANNProof across all settings, with speedups of 1.11--2.47$\times$. However, VIDA*-Q is slower than ANNProof-Q at all four recall targets for $k=10$ and $25$, and at targets 0.75 and 0.80 for $k=50$. For example, at $k=10$ and target recall 0.75, ANNProof-Q takes only 118.0 ms, compared with 242.4 ms for VIDA*-Q. As $k$ and the target recall increase, the latency gap between VIDA*-Q and ANNProof-Q narrows, with VIDA*-Q becoming slightly
faster at $k=50$ and target recalls 0.85 and 0.90. Under low selectivity or low recall, reduced search effort makes raw-vector transmission less costly, especially when quantization reduces the number of candidates requiring exact-distance verification. Under these lighter workloads, the communication savings from compact proofs may not offset proof generation and verification costs.
\fi
\end{document}
\endinput